\documentclass[pra,twocolumn,superscriptaddress,showpacs,amsmath,amstex,amssymb,citeautoscript]{revtex4-1}

\usepackage[T1]{fontenc}
\usepackage[utf8]{inputenc}
\usepackage{lipsum}
\usepackage{amsmath}
\usepackage{amssymb}
\usepackage{bbm}
\usepackage{braket}
\usepackage{xcolor}
\usepackage{pifont}
\usepackage[normalem]{ulem}
\usepackage[mathscr]{euscript}
\usepackage[shortlabels]{enumitem}
\usepackage[justification=justified]{subcaption}
\usepackage{graphicx}
\usepackage{lipsum}
\allowdisplaybreaks
\usepackage{float}
\usepackage{graphicx}
\usepackage{dsfont}
\usepackage{comment}
\usepackage[colorlinks=true]{hyperref}  
\hypersetup{
    bookmarks=true,         % show bookmarks bar?
    unicode=false,          % non-Latin characters 
    pdftoolbar=true,        % show Acrobat
    pdfmenubar=true,        % show Acrobat 
    pdffitwindow=false,     % window fit to page when opened
    pdfstartview={FitH},    % fits the width of the page to the window
    pdftitle={Quantum Twisting Microscope and Quantum Geometry},    % title
    pdfauthor={Pupim and Scheurer},     % author
    pdfsubject={},   % subject of the document
    pdfcreator={},   % creator of the document
    pdfproducer={}, % producer of the document
    pdfkeywords={} {} {}, % list of keywords
    pdfnewwindow=true,      % links in new window
    colorlinks=true,       % false: boxed links; true: colored links
    linkcolor=blue, %red,          % color of internal links (change box color with linkbordercolor)
    citecolor=blue,        % color of links to bibliography
    filecolor=magenta,      % color of file links
    urlcolor=blue           % color of external links
}

\newcommand{\equref}[1]{Eq.~(\ref{#1})}
\newcommand{\equsref}[2]{Eqs.~(\ref{#1}) and (\ref{#2})}
\newcommand{\secref}[1]{Sec.~\ref{#1}}
\newcommand{\figref}[1]{Fig.~\ref{#1}}
\newcommand{\refcite}[1]{Ref.~\onlinecite{#1}} 
\newcommand{\refscite}[1]{Refs.~\onlinecite{#1}}

\newcommand{\appref}[1]{Appendix~\ref{#1}}

\newcommand{\sign}{\,\text{sign}}
\renewcommand{\approx}{\simeq}

\renewcommand{\vec}[1]{\boldsymbol{#1}}

\definecolor{wrongultramarine}{rgb}{1,0.5,0}

\begin{document}

\title{Quantum twisting microscopy as a momentum-resolved probe of \\ quantum geometry and sublattice symmetries}

\author{Lucas V.~Pupim}
\affiliation{Institute for Theoretical Physics III, University of Stuttgart, 70550 Stuttgart, Germany}

\author{Mathias S.~Scheurer}
\affiliation{Institute for Theoretical Physics III, University of Stuttgart, 70550 Stuttgart, Germany}

\begin{abstract}
In two-dimensional (2D) flat-band systems, as engineered, for instance, by stacking and twisting van der Waals materials, the momentum dependence of the Bloch states is a key factor in determining the interaction-induced emergence of different phases. Hence, methods of probing this form of ``quantum geometry’’ are crucial to understanding these systems. Yet, extracting the momentum dependence of the underlying quantum geometric tensor from measurements is challenging in 2D.
Here, we propose employing a quantum twisting microscope (QTM) with a sublattice-polarized tip to fill this gap. We show how expectation values of sublattice Pauli matrices can be probed across the scanned momentum path. This allows for the direct measurement of crucial symmetries characterizing the different candidate instabilities in graphene-based moiré systems. We furthermore show that the quantum metric component along the scan direction can also be extracted; the full quantum geometric tensor is also accessible, given that more than one sublattice polarization can be used. Our work illustrates the potential and extends the versatility of the QTM approach. If implemented in future experiments, this would provide unprecedented access to the symmetry signatures of the correlated phases and to the underlying quantum geometry in twisted bilayer graphene and related systems.
\end{abstract}

\maketitle

\section{Introduction}

While non-trivial electronic band energies, such as van Hove singularities, are well-established driving forces of rich many-body physics, the role of complex momentum-dependencies of the Bloch wave functions has recently also attracted a lot of attention in the condensed matter community \cite{yu2025quantum,Verma2026Apr,f78t-ky69}. In analogy to the geometric phase under adiabatic variation of the continuous parameter of time, crystalline momentum $\vec{k}$ is also a continuous parameter (in the thermodynamic limit) such that the Bloch states $\ket{\psi_{\vec{k}}}$ encode a geometric structure, too. This is most commonly quantified, to leading order, by the quantum geometric tensor,
\begin{equation}
Q_{ij}(\vec{k}) = \braket{\partial_{k_i}\psi_{\vec{k}}|(\mathbbm{1} - \ket{\psi_{\vec{k}}}\bra{\psi_{\vec{k}}})|\partial_{k_j}\psi_{\vec{k}}} = Q^*_{ji}(\vec{k}).  \label{QGT}
\end{equation}
For instance, for two-dimensional (2D) systems, we have $i,j=x,y$ and $\text{Im}\,Q$ has only one independent component---the Berry curvature $\Omega = -2\text{Im} \,Q_{xy}$---and three independent real parts, defining the ``quantum metric’’ $g_{ij} = \text{Re} \,Q_{ij}$. In the last few years, the relevance of these quantum geometric effects \cite{yu2025quantum,Verma2026Apr} has been recognized to extend far beyond just the Berry curvature and Chern numbers. For instance, it was shown to have crucial consequences for transport \cite{2024arXiv241219056C,PhysRevResearch.4.013001,PhysRevB.110.174423,PhysRevB.108.155108,PhysRevB.105.085154,PhysRevB.102.165151,xkxw-1134}, the stiffness \cite{ReviewQuantumGeometry,DisorderSFW,2025arXiv251005224K}, the superconducting response at finite momenta \cite{PhysRevB.108.094508}, robustness of flat-band superconductivity~\cite{PhysRevB.111.L020506}, Cooper pair size \cite{PhysRevB.111.014502}, spectral weight transfer \cite{PhysRevB.104.L100501}, bound-state localization \cite{2025arXiv250515891L}, for pairing glues \cite{PhysRevLett.134.176001,JahinLin,QGDiscrepancy,2025arXiv250913407S,StandfordWithTheRelativeChirality,ChristosEnergetics}, and disorder sensitivity \cite{2025arXiv251019943S}. Although (certain components of) $Q_{ij}$ enter in many physical quantities, directly extracting it from measurement, in particular, its momentum dependence is challenging since many probes are not momentum resolved. Besides other promising directions \cite{gianfrate2020measurement,nzc1-wg6m,PhysRevB.111.125161,Sala2025Aug,yu2020experimental,tan2019experimental}, photoemission is a particularly powerful route towards momentum-dependent access \cite{kim2025direct,Kang2025Jan,lange2026estimating} but remains challenging in 2D systems.

Meanwhile, twisted van der Waals stacks \cite{Andrei2020Dec,Sun2024Feb} have been established as a flexible playground for the interplay of quantum geometry and correlation effects in 2D. For instance, in graphene-based moiré superlattices, such as twisted bilayer graphene (TBG), the non-trivial quantum geometry \cite{PhysRevLett.123.237002, PhysRevB.101.060505} is intimately linked with the two sublattice degrees of the constituent honeycomb lattices. The presence of strong interactions is known to give rise to a variety of different forms of spontaneous symmetry breaking throughout the phase diagram of these systems, which in turn is believed to be crucially shaped by the non-trivial momentum dependence of the systems’ Bloch states. This is why extracting $Q$ in TBG and related systems from experiment is actively pursued \cite{krishna2025terahertz,PhysRevResearch.4.013164,tanaka2025superfluid}, but providing momentum-resolved access remains an open challenge.

A very recently developed experimental technique that is capable of resolving the momentum-dependence of the band energies in 2D materials is the quantum twisting microscope (QTM) \cite{inbar2023quantum,2026arXiv260403483B,xiao2026imaging,lee2026revealing,birkbeck2025quantum}. As illustrated schematically in \figref{fig:1}(a), it is based on a planar and thus momentum-conserving tunnel junction between a graphene ``tip’’ and the sample of interest, e.g., single-layer graphene or TBG. Importantly, the twist angle $\theta$ between the two can be varied \textit{in situ}. As the tip’s Dirac cones trace out paths in the sample Brillouin zone when varying $\theta$, the bias-voltage and $\theta$ dependence of the tunneling current maps out the sample’s energy bands on those paths \cite{inbar2023quantum,wei2025dirac}. Apart from electronic spectral functions, also phonons \cite{birkbeck2025quantum, PhysRevB.110.205407} have been probed with QTM and theoretical works have proposed using QTM to study plasmons \cite{v8zx-1vhr}, superconductivity \cite{62rf-8m43,qsx1-7zmy}, fractionalization \cite{ PhysRevB.109.035127,PhysRevB.111.075108, anyons}, and magnetism \cite{PhysRevB.110.045116}.

In this work, we theoretically study how the QTM approach can be extended to extract the quantum geometric tensor along the scanned momentum paths in the sample. To achieve this, we propose using a sublattice-polarized graphene tip, which could, for instance, be achieved by near alignment of the graphene layer with the hexagonal boron nitride (hBN) gate dielectric \cite{wang2016gaps,PhysRevB.76.073103,hunt2013massive,chen2014observation,PhysRevB.84.195414,jung2017moire,woods2014commensurate}. As shown in \figref{fig:1}(b), this gaps out the Dirac cone; recognizing that the tunneling currents at the voltages corresponding to the band edges of the tip involve matrix elements with the (nearly) sublattice polarized tip wave function, we obtain sublattice-resolved information about the system. We show that this not only allows to directly probe whether key symmetries are preserved or broken (with momentum resolution) but also provides access to certain components of $Q_{ij}(\vec{k})$. More specifically, with a single value of the sublattice polarization $\Delta$, one can extract the components of $Q$ along the scanned path, while access to two different $\Delta$ even also allows to reconstruct the remaining components.

\section{Results}
\subsection{QTM with sublattice polarized tip}
To be self-contained, define the notation, and extend previous works on QTM to sublattice polarized tips, we here describe the overall setup in detail and how the tunneling current is computed. While the basic strategy is more generally applicable, we will, however, focus on graphene-based systems for concreteness. As such, both the tip and sample are based on graphene layers or, more generally speaking, have the same symmetries as graphene. 
More explicitly, the tip is taken to be a stack of graphene and hBN as is commonly done in QTM experiments \cite{PhysRevB.110.205407,inbar2023quantum,lee2026revealing}; however, we assume here that the two are within a few degrees of being aligned. This is known to induce a $4-50$meV sublattice gap $\Delta$ \cite{wang2016gaps,PhysRevB.76.073103,hunt2013massive,chen2014observation,PhysRevB.84.195414,jung2017moire,woods2014commensurate} around graphene's primary Dirac points at charge neutrality, which can be tuned by the twist angle between the two lattices. While the associated moiré superlattice between hBN and graphene in the tip does create an additional reconstruction of the bands with secondary Dirac cones at other energies \cite{Ponomarenko2013May,PhysRevB.84.195414,jung2017moire}, we will first focus on energies close to charge neutrality. Furthermore, the moiré periodicity is smaller than 14 nm \cite{jung2017moire} near alignment. This means that the planar QTM tip can incorporate enough unit cells \cite{inbar2023quantum} and be considered periodic. Then the Hamiltonian for the tip, with momentum deviations $\vec{k}=(k_x,k_y)^T$ from the original K point of the graphene tip ($\vec{K}_T$) being much smaller than the linear mini Brillouin zone size of the hBN-graphene superlattice, can be written as $h_T(\vec{k}) =  v_D( k_x  \sigma_x + k_y \sigma_y ) + \Delta \sigma_z -\mu_T$; here, $\mu_T$ is the chemical potential of the tip and the set of Pauli matrices $\sigma_i$, $i=0,x,y,z$, act in sublattice space. Throughout this work, we will suppress the spin quantum numbers. The bands of this Hamiltonian are given by $\xi^T_{\vec{k},\lambda} = \lambda \sqrt{( v_D \vec{k})^2+ \Delta^2} - \mu_T$ with eigenvectors
\begin{equation}
\label{tip_states_general}
    \zeta_{\vec{k},\lambda} = \frac{1}{\sqrt{2}} \begin{pmatrix} \sqrt{ 1+ \lambda \cos \alpha_k} \\
    \lambda e^{i \varphi_{\vec{k}}} \sqrt{1 - \lambda \cos \alpha_k}
    \end{pmatrix}.
\end{equation}
Here $\varphi_{\vec{k}}$ is the polar angle of $\vec{k}$, $\tan \alpha_k = \frac{ v_D k}{\Delta}$ with $k=|\vec{k}|$, and $\lambda= \pm$ indexes the band. The Hamiltonian, bands, and Bloch functions associated with the K$'$ point of graphene are obtained from those above by time reversal.

\begin{figure}[tb]
    \centering
    \includegraphics[width=\linewidth]{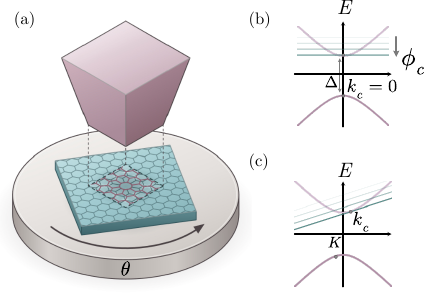}
    \caption{(a) Schematic of the quantum twisting microscope. The tip has a flat contact area with a 2D material compatible with the sample's lattice. A given twist angle $\theta$ defines the allowed tunneling momenta and through this mechanism the junction works as a momentum-resolved probe. We propose to use a non-zero sublattice polarization $\Delta$ in the tip to retrieve the quantum metric and probe symmetries. (b) and (c) represent the energetic condition $\phi_c^\lambda$ to have momentum resolution of the sample. If the sample's band is flat (b), the momentum probed corresponds to the tip's (rotated) Dirac point $K$, otherwise the probed momentum by the tip's conductance band is shifted by $\vec{k}_c$ (or $-\vec{k}_c$ for the valence band).}
    \label{fig:1}
\end{figure}

Let us denote the dispersion and the (normalized) Bloch wave functions of the band of interest in the sample by $\xi^S_{\vec{k}'}$ and $\ket{\Psi^S_{\vec{k}'}}$, respectively, with $\vec{k}'$ in the Brillouin zone of the sample. Since we focus on graphene-based systems, we can further split $\ket{\Psi_S} = (\ket{\psi_{A}},\ket{\psi_{B}})^T$ into its sublattice A and B parts for future reference.
Note that the sample could have multiple layers. In that case, $\ket{\Psi^S_{\vec{k}'}}$ will denote the projection of the Bloch band onto the topmost layer, since the electrons will primarily tunnel into the layer closest to the tip. In our convention, $\ket{\Psi^S_{\vec{k}'}}$ is already renormalized, by a factor $1/\sqrt{p_S(\vec{k})}$, to have norm $1$ after projection, $\braket{\Psi^S|\Psi^S} = 1$. For instance, in TBG with its two layers (related by a $C_{2x}$ rotational symmetry), the top-layer weight $p_S$ is approximately independent of momentum and given by $p_S(\vec{k})\approx 0.5$ \cite{persky2026optical}; in fact, since tunneling preserves crystalline momentum, the tunnel matrix elements further entail a projection to a single momentum-lattice site. We have checked that $p_S(\vec{k})\approx \text{const.}$ is still approximately valid on the relevant momentum path, in particular, around the magic angle, see \appref{ProjectionsInTBG}. 

Now, we turn to the computation of the current flowing between the sample and tip, which is the key observable studied in QTM experiments. We follow similar steps as in \refscite{inbar2023quantum,wei2025dirac}. For a given bias $eV$ between the sample and tip, ``Fermi's golden rule'' yields the tunnel current
\begin{align}\begin{split}
    I = 2\pi e &\sum_{\lambda}\sum_{\vec{k},\vec{Q}}
\bigl\{
\left[f\left(\xi^S_{\vec{p}}\right)-f\left(\xi^T_{\vec{p},\lambda}\right)\right] \\ 
 & \quad\times \left|T_{\lambda}(\vec p)\right|^2
\delta\left(\xi^S_{\vec p }-\xi^T_{\vec p ,\lambda}+eV\right)
\bigr\}\bigr|_{\vec p = \vec k+\vec Q}. \label{eq:current-1}
\end{split}\end{align}

Here we assumed that the spectral function of the sample and tip are simply Dirac deltas, $\vec{p} \in \mathbb{R}^2$ is defined in the extended Brillouin zone scheme with $\xi^S_{\vec p }$ and $\xi^T_{\vec p ,\lambda}$ extended to be periodic with respect to their underlying reciprocal lattices. Furthermore, $f$ are the Fermi functions and $T_{\lambda}(\vec{p}) = \sqrt{p_S}\langle \zeta_{\vec{p},\lambda}  | \hat{T} (\vec{p}) | \Psi^S_{\vec{p}} \rangle$ are the tunneling amplitudes between the sample's and the tip's Bloch bands, where $\hat{T} (\vec{p})$ is the tunneling operator to be specified below. The additional factor $\sqrt{p_S}$ comes from the renormalization of the sample Bloch states. The sum over $\vec{Q}$ in \equref{eq:current-1} accounts for momentum conservation up to a reciprocal lattice vector, i.e., Umklapp tunneling, while $\vec{k}$ is defined in the first Brillouin zone. One could choose this sum to be either over sample or tip vectors, since one uniquely defines the other \footnote{By virtue of momentum conservation, $\vec{k} + \vec{Q} = \vec{k}' + \vec{Q}'$, across the interface; here $\vec{Q}$ ($\vec{Q}'$) are the reciprocal lattice vectors of the tip (sample).}. We also note that $T_\lambda(\vec{p)}$ typically decay fast for $|\vec{p}|$ larger than the interlayer separation \cite{bistritzer2010moire}, which will allow us to truncate the summation over the reciprocal vectors $\vec{Q}$. We will discuss this truncation and the properties of $T_{\lambda}$ later on.

For the moment, let us further simplify the current expression in \equref{eq:current-1} in a way that will allow us to extract the components of the wave function and quantum metric. We  approximate the summations in $\vec{k}, \vec{G}$ as an integral and write \cite{wei2025dirac}
\begin{equation}\begin{split}
\label{eq:current-exact}
I=\frac{e \Omega}{2 \pi }  \sum_{ \lambda} \int_{\delta \epsilon_{\vec{p}}^{\lambda}=\phi} \frac{d p_{\|}}{\left|\vec{\nabla}_{\vec{p}} \delta \epsilon_{\vec{p}}^{\lambda }\right|}
\left[f\left(\xi_{\vec{p} }^S\right)-f\left(\xi_{\vec{p}, \lambda}^T\right)\right] \left|T_{\lambda }(\vec{p})\right|^2,
\end{split}
\end{equation}
where $\delta\epsilon^{\lambda}_\mathbf{p} = \xi^S_{\vec{p}} + \mu_S -\xi^T_{\vec{p},\lambda} - \mu_T$ and $\phi = -eV -\mu_T + \mu_S$ is the electrostatic potential difference. Here $\Omega$ is the junction area. Note that a given $\phi(eV,\mu_S, \mu_T)$ defines the contour of integration, see \figref{fig:1}(b) and (c). More precisely, since $\vec{p} \in \mathbb{R}^2$, it defines an infinite set $\{C^\lambda_n(\phi)\}$ of disconnected contours, each in the vicinity of a Dirac point and associated with either the conduction ($\lambda=+$) or valence band ($\lambda=-$).

If these contours shrink to isolated momentum points $\vec{p}=\vec{p}^\lambda_{c,n}$ at a critical $\phi=\phi_c^\lambda$, the transport properties will be associated with those discrete momenta. Together with the exponential decay of $T_\lambda(\vec{p})$, which allows us to focus on only a small subset of $\vec{p}=\vec{p}^\lambda_{c,n}$ that are typically further related by symmetry, one effectively obtains momentum-resolved information about the system. For instance, in our case of a graphene tip, one only needs to take into account (at given $\lambda$ and valley) the three $\vec{p}^\lambda_{c,n}$ that correspond to the same Dirac cone in graphene's Brillouin zone---in analogy to the frequently used approximation in the continuum model of TBG \cite{dos2007graphene,MeleModel,bistritzer2010moire,balents2019general}; one therefore only has three ($C_{3z}$-related) $\vec{p}^\lambda_{c,n}$.

For notational simplicity, let us temporarily focus on only one such momentum point, which we label as $\vec{p}^\lambda_{c,1} = \vec{k}^\lambda_{c}$; the others are given by $\vec{p}^\lambda_{c,n} = \vec{k}^\lambda_{c} + \vec{Q}_n$ with reciprocal lattice vectors $\vec{Q}_{n>1} \neq 0$. Using the gapped Dirac model $h_T(\vec{k})$ for the tip, the critical condition for $\phi$ that shrinks the contour $\delta \epsilon_{\vec{p}}^{\lambda}=\phi$ to an isolated point in band $\lambda=\pm$ [cf.~\figref{fig:1}(c) for the upper band, $\lambda=+$] and the associated momentum read as 
\begin{equation}
     \phi_c^\lambda= \xi^S_{0} + \mu_S -\frac{\lambda|\Delta|}{\gamma} \,\,\text{ and }\,\, \vec{k}^\lambda_c = \frac{\lambda |\Delta| \gamma} {v_D^2} \vec{v}_0, \label{CricialValueofPhi}
\end{equation}
respectively (see \appref{app:contour} for details). We defined $\gamma = (1 - \vec{v}_0^2/v_D^2)^{-1/2}$, and expanded the sample dispersion around the position (in our convention $\vec{k}=0$) of the tip's Dirac cone, $\xi^S_{\vec{k}} \sim \xi^S_{0} + \vec{v}_{0} \cdot \vec{k} + \frac{1}{2}\vec{k}  M^{-1}\,\vec{k}$. Here $M$ is the mass tensor with $M_{ij}^{-1} = \frac{\partial^2 \xi^S_{\vec{k}}}{\partial_{k_i}\partial_{k_j}}\big|_{\vec{k}=\vec{k}_c^\lambda}$. Before proceeding, a few comments are in order. First, not only $\phi_c^\lambda$ but also $\vec{k}^\lambda_c$ depends crucially on the band index $\lambda$. Second, the expansion used for $\xi^S_{\vec{k}}$ only needs to be valid on the small scale of $\vec{k}^\lambda_c$ and is thus expected to be a good approximation for generic sample momenta. In fact, when $\vec{v}_0 \neq 0$ the curvature correction is expected to be negligible and we did already neglect it in \equref{CricialValueofPhi}. However, we noticed empirically in our examples that keeping it as a correction in the density of states below yields more accurate results when $\vec{v}_0$ becomes small. Third, if either the gap in the tip vanishes, $\Delta=0$, or the sample's band is perfectly flat $\vec{v}_0 = 0$ [cf.~\figref{fig:1}(b)], then $\vec{k}^\lambda_c=0$ which corresponds to the Dirac point itself, e.g., the K point, and the expansion used above becomes exact. Otherwise, we have a finite momentum shift $\vec{k}^\lambda_c$ along $\vec{v}_0$ for the tip's valence band and in the opposite direction for the conduction band, since $\vec{k}^\lambda_c$ is proportional $\lambda$.

Consequently, for voltages $V$ close to $V^\lambda_c$, defined via $\phi_c^\lambda= -eV^\lambda_c -\mu_T + \mu_S$ and  such that the contours in \equref{eq:current-exact} shrink to $\vec{p}^\lambda_{c,n}$, we can approximate $\vec{p} \approx \vec{p}^\lambda_{c,n}$ in the Fermi functions and write
\begin{equation}
\label{eq:current_pc}
I\approx 2 \pi e \Omega  \sum_{n,\lambda} \left[f\left(\xi_{\vec{p}^\lambda_{c,n} }^S\right)-f\left(\xi_{\vec{p}^\lambda_{c,n} \lambda}^T\right)\right] \left\langle\left|T_{\lambda,n}\right|^2 \right\rangle_{\hspace{-0.2em}\phi} \nu^\lambda_{c,n}(\phi),
\end{equation}
with the density of states 
\begin{equation}
\label{eq:JDOS}
\nu^\lambda_{c,n}(\phi) \equiv \frac{1}{(2 \pi)^2} \int_{C^\lambda_n(\phi)} \frac{d p_{\|}}{\left|\vec{\nabla}_{\vec{p}} \delta \epsilon_{\vec{p}}^{\lambda }\right|}
\end{equation}
associated with the contour $C_n^\lambda(\phi)$. Here, we also defined
\begin{equation}
    \left\langle\left|T_{\lambda,n}\right|^2 \right\rangle_{\phi} =  \frac{1}{(2 \pi)^2  \nu_{c,n}^\lambda (\phi) } \int_{C^\lambda_n(\phi)} \frac{d p_{\|}}{\left|\vec{\nabla}_{\vec{p}} \delta \epsilon_{\vec{p}}^{\lambda }\right|} |T_{\lambda}(\vec{p})|^2.
\end{equation}
In our specific case of a gapped Dirac cone in the tip and the expanded sample dispersion, one finds (see \appref{app:JDOS} for derivation)
\begin{equation}
\label{eq:nu_c}
\begin{split}
\nu^\lambda_c = \frac{ \Theta(\lambda [\phi_c^\lambda-\phi]) }{2 \pi v_D^2} \gamma^2 \left[
\gamma|\phi_c^\lambda - \phi|\,(1+3\lambda\kappa)\right.\\
\left. + |\Delta|\,(1+\lambda\kappa)\right],                 
\end{split}\end{equation}
where $\kappa = \frac{\Delta \gamma}{2 v_D^2} \,\text{Tr}[M^{-1}]$ is the mean curvature of the sample relative to the tip curvature.
Intuitively, the first (second) term in \equref{eq:nu_c} is reminiscent of the linear growth of the density of states of a Dirac cone (of the constant density of states of a two-dimensional quadratic band) and dominates for large $|\phi|$ (large $|\Delta|$); the Heaviside step function $\Theta$ captures the vanishing contributions inside the gap of the tip band structure. 

\begin{widetext}
Furthermore, the factor that appears in the current expression \equref{eq:current_pc} becomes 
\begin{equation}
\label{eq:T_nu}
\left\langle\left|T_{\lambda,n}\right|^2 \right\rangle_{\hspace{-0.2em}\phi}\nu^\lambda_{c,n}(\phi) = \frac{\Theta\big(\lambda(\phi_{c}^\lambda-\phi)\big)}{2\pi v_D^2}\Big[
|\Delta|(1+\lambda\kappa)\,|T_{\lambda}(\vec{p}_{c,n}^\lambda)|^2
+ |\phi_{c}^\lambda-\phi|\left((1+3\lambda\kappa)\,|\bar T_{n,\lambda}|^2
+ \lambda\kappa\,|\delta T_{n,\lambda}|^2\right)
\Big] + \mathcal{O}\left(\beta^2\right)
\end{equation}
with $\beta=|\vec{v}_0|/v_D$, $ |\overline{T}_{n,\lambda}|^2 = (|T_+(\vec{p}^\lambda_{c,n})|^2+|T_-(\vec{p}^\lambda_{c,n})|^2)/2$, and $|\delta T_{n,\lambda}|^2=(|T_+(\vec{p}^\lambda_{c,n})|^2-|T_-(\vec{p}^\lambda_{c,n})|^2)/2$. Here, we expressed these two terms using $T_\pm(\vec{p}^\lambda_{c,n})$ for convenience, but this is just a result of the tip's wavefunctions deviating from their form right at $\vec{p}^\lambda_{c,n}$ when $\phi\neq \phi_c^\lambda$ rather than a contribution of both tip bands at a given energy, see \appref{app:expansion}. 
\end{widetext}

For our goal, it is important to note that the conductance $dI/dV$ becomes proportional to $\frac{d}{dV}\left|T_{\lambda,n}\right|^2\,\nu^\lambda_{c,n}$ which, as a result of the form of $\left|T_{\lambda,n}\right|^2\,\nu^\lambda_{c,n}$ in \equref{eq:nu_c}, is proportional to a step function plus a Dirac delta. While the step function part involves $\left|T_+(\vec{p}_{c,n})\right|^2 \pm \left|T_-(\vec{p}_{c,n})\right|^2$, the Dirac delta term that only exists for finite $\Delta$ is proportional to the sublattice polarized tunneling elements and therefore sublattice resolved. While both terms provide useful information, we will later first focus on this latter term to extract sublattice symmetries and the quantum geometry of the sample.
The conductance derivative has an even sharper profile with the sum of a Dirac delta and its derivative. As a byproduct, in an experimental realization, this means that our proposed method with a gapped tip should have a sharper signature and therefore more resolution than a gapless QTM tip.

Changing the twist angle $\theta$ between the tip and the sample moves the scanned momenta $\vec{p}_{c,n}$ relative to the sample's Brillouin zone and effectively provides momentum-resolved information on those paths. We reiterate our conventions where $\vec{k}=0$ refers to one of the K points of the tip graphene layer, while the sample Hamiltonian is rotated when tuning $\theta$. The paths in the Brillouin zone of the sample are equivalent if the $\vec{p}_{c,n}$ are mapped into a single momentum through symmetry, as is the case for (undistorted) graphene layers on both sides. Further note that twisting away from different commensurate angles gives different scanned paths, $cf.$ \refcite{wei2025dirac}. 

\subsection{Sublattice expectation values in flat-band limit}
To address our ultimate goal of extracting components of the quantum geometric tensor, let us first take the flat-band limit in the sample, $\vec{v}_0 \rightarrow 0$.  Consequently, we get $\vec{k}_c=0$ and the aforementioned three momenta $\vec{p}^\lambda_{c,n}$ to keep in \equref{eq:current_pc} are simply given by $\vec{p}^\lambda_{c,n} = \vec{Q}_n$, with $\vec{Q}_1 = 0$, $\vec{G}_2=\frac{4 \pi}{3 a}(\sqrt{3}/2, 1/2)$, and $\vec{G}_3=\frac{4 \pi}{3 a}(\sqrt{3}/2,- 1/2)$ where $a$ is the distance between the carbon atoms. As $\vec{p}^\lambda_{c,n}$ are now independent of $\lambda$, we will temporarily drop the superscript. The associated matrix elements of the tunneling operator $\hat{T}$ for two graphene layers are well known from the continuum model of TBG \cite{dos2007graphene,MeleModel,bistritzer2010moire,balents2019general} and given by \begin{equation}
\label{eq:tunn_matrix}
    \hat{T} (\vec{p}_{c,n})=  \begin{pmatrix} w_0\,e^{i\chi} & w_1\,e^{-i\vartheta_n } \\[4pt] w_1\,e^{i\vartheta_n } & w_0\,e^{-i\chi} \end{pmatrix},
\end{equation}
with $\vartheta_n = \tfrac{2\pi(n-1)}{3}$ and the matrix acting in sublattice space.

The flat-band limit further simplifies the form of $T_\lambda(\vec{p}_{c,n})$ since the tip states in \equref{tip_states_general} are evaluated at $\vec{k}=0$ and, thus, fully polarized, i.e., $\zeta_{+} = (1, 0)^T$ and $\zeta_{-} = (0, 1)^T$. One finds from \equref{eq:tunn_matrix}
\begin{equation}
\begin{split}
    \frac{|T_{+}(\vec{p}_{c,n})|^2}{p_S} &= w_0^2\,|\psi_A|^2 + w_1^2\,|\psi_B|^2
   \\ &\quad + 2 w_0 w_1\,\mathrm{Re}\!\left[e^{i(\chi+\vartheta_n)}\,\psi_A\psi_B^*\right], \label{TunnelComponents}
    \end{split}
\end{equation}
and the tunneling elements with the valence band state $|T_{-}(\vec{p}_{c,n})|^2$ can be obtained by swapping $w_0 \leftrightarrow w_1$. Note that the presence of the almost aliged hBN  in principle lowers the symmetries in graphene which, to higher order, would lead to additional corrections to the tunneling terms; we will, however, neglect these corrections here as is also commonly done in multilayer continuum-model calculations.

Note that, by construction, if the sample has $C_{3z}$ symmetry, all three tunneling terms will be equivalent, $|T_{\lambda}(\vec{p}_{c,n})|^2 = |T_{\lambda}(\vec{p}_{c,1})|^2$ \footnote{Mathematically, this comes from the sublattice-dependent phase acquired during the sample's rotation canceling with $\vartheta_n$ in $\text{Re}\left[e^{i(\chi+\vartheta_n)}\,\psi_A\psi_B^*\right]$ in \equref{TunnelComponents}.}. Since the sample's dispersion would then also be symmetric under $C_{3z}$, the three channels, $n=1,2,3$ in \equref{eq:current_pc}, all yield the same contribution, and the current gains a factor of 3. Furthermore, valley and spin degeneracy, if present, will give an extra factor $(2\times 2)$. 

For our goal of probing the quantum metric, the important bit is to notice that the term proportional to $|\Delta|$ in $dI/dV$ (and its derivative) is proportional to $|T_{\lambda}(\vec{p}_c)|^2$ and that we can rewrite that part of the conductance measurement (or alternatively of its derivative) for $V$ close to $V_c^\lambda$ as
\begin{equation}
\label{eq:gamma_def}
   \Gamma_{\lambda} = \mathcal{N} p_S \braket{\Psi^S _{\vec{p}_{c,1}}| \hat{P}_{\lambda} |  \Psi^S_{\vec{p}_{c,1}}}.
\end{equation}
The proportionality constant $\mathcal{N}$ absorbs the prefactor coming from the different flavors and the three different $n$. In addition, we defined $\hat{P}_{\pm} = \hat{T}^\dagger (\vec{p}_{c,1})\ket{\zeta_{\pm}}\bra{\zeta_{\pm}}\hat{T} (\vec{p}_{c,1}) \equiv \ket{p_{\pm}}\bra{p_{\pm}}$ with $\ket{p_+} = (w_0 e^{-i\chi},w_1)^T$ and $\ket{p_{-}}= (w_1,w_0e^{i\chi})^T$. Therefore, we can interpret our junction as a probe for the expectation values of the projectors $\hat{P}_{\pm}$ on the sample's (top layer) state $\Psi^S(\vec{p}_{c,1})$.

These projectors can be cast as $\hat{P}_{\pm} = \vec{d}_{\pm} \cdot \vec{\sigma}\, + \frac{w_0^2 + w_1^2}{2} \sigma_0 $, where $\vec{\sigma}=(\sigma_x,\sigma_y,\sigma_z)^T$ is the vector of Pauli matrices in sublattice space and 
\begin{equation}
    \vec{d}_{\pm} = (w_0 w_1 \cos \chi, w_0 w_1 \sin \chi, \pm \frac{w_0^2-w_1^2}{2}).
\end{equation}
One can see that linear combinations of the tunneling measurements (and therefore of the projectors) provide access to the expectation value of operators in a 2 dimensional plane (out of the 4 dimensional operator space). Note that if $w_0 = w_1$, the projectors are the same and this 2D plane collapses into a line. However, this is a fine-tuned point since $w_0$ and $w_1$ are not related by symmetry \cite{balents2019general} and thus generically not identical. For instance, for small twist angles, $w_1/w_0 \approx 0.8 $ has been estimated \cite{li2024evolution}, which is what we will use here in our examples. We emphasize that QTM can even be used to extract $w_0/w_1$ from measurements \cite{wei2025dirac}, which could supplement our proposed protocol here to obtain a sample-specific determination of the tunnel matrix elements in \equref{eq:tunn_matrix}. 
As it will turn out to be convenient, let us employ the orthogonal pair $\hat{P}_{1,2} = \frac{1}{2}(\hat{P}_+ \pm \hat{P}_-)$, explicitly given by
\begin{equation}
\begin{split}
    \hat{P}_1 &= \frac{w_0^2 + w_1^2}{2}\sigma_0+ w_0 w_1 ( \cos \chi,  \sin \chi,0) \cdot \vec{\sigma},\\
    \hat{P}_2  &= \frac{w_0^2 - w_1^2}{2} \sigma_z \label{ProjectorsWeHaveAccessTo}
\end{split}   
    \end{equation}
and associated with the measurements $\Gamma_{1,2} =\frac{1}{2} (\Gamma_{+} \pm \Gamma_-)$. 
If one restricts themselves to linear combinations, this plane is the extent of what can be probed. Defining the expectation values $n_i(\vec{k}') := \langle \Psi^S_{\vec{k}'} |\sigma_i| \Psi^S_{\vec{k}'} \rangle$, we can extract $n_z$ and $n_{\chi} =  \left(\cos \chi \,\, n_x +\sin \chi \,\, n_y  \right)$ directly from the measurements $\Gamma_{1,2}$ via 
\begin{equation}
    n_\chi = \frac{2\Gamma_1 - p_S\mathcal{N}(w_0^2+w_1^2)}{2 p_S\mathcal{N} w_0 w_1}, \,
    n_z = \frac{2 \Gamma_2}{p_S\mathcal{N}(w_0^2 - w_1^2)}. \label{nChiNz}
\end{equation}
While this is not yet enough for the quantum geometric tensor, it does provide crucial symmetry-sensitive insights into the sample's wavefunctions in sublattice space. For instance, the product $C_{2z}\Theta$ of time-reversal, $\Theta$, and two-fold rotational symmetry, $C_{2z}$, is central to graphene-based heterostructures, such as TBG, and their correlated physics, where this symmetry can be broken spontaneously. It is represented as $\sigma_x \mathcal{K}$, where $\mathcal{K}$ is complex conjugation, such that $n_z \neq 0$ if and only if it is broken, proving momentum-resolved access to the degree to which this symmetry is preserved.

Another important yet approximate symmetry of graphene moiré systems is chiral symmetry $C$ \cite{PhysRevLett.122.106405}, which relates the upper and lower flat bands. It is given by $C=\sigma_z$ and, thus, requires that $n_\chi$ ($n_z$) have opposite (the same) values in the two flat bands. As with $C_{2z}\Theta$ above, this can be used to characterize interaction-induced symmetry breaking. Furthermore, $C$ is essential for obtaining (nearly) ideal Chern bands \cite{PhysRevLett.122.106405,PhysRevResearch.3.023155}, which in turn is relevant, e.g., for the emergence of fractional Chern insulators \cite{PhysRevB.108.205144}. As such, our proposed technique would allow non-trivial, momentum-resolved experimental access to the nature of the TBG wave function.

\subsection{Quantum metric}
However, to obtain $Q_{ij}$, we need to go beyond linear combinations of the operators in \equref{ProjectorsWeHaveAccessTo} since it requires all three components of $\vec{n}$ independently. 
As tuning the twist angle $\theta$ between the sample and the tip effectively changes the momentum at which $\vec{n}$ is evaluated, we can write the quantum metric component along the scanned path as
\begin{equation}
    g_{\theta \theta} = \frac{1}{4K^2} |\partial_\theta \vec{n}|^2, \label{gthetatheta}
\end{equation}
where $K$ is the length of the K point in graphene. Our two measurements $\Gamma_{1,2}$ give us two, $n_{\chi}$ and $n_z$, out of the three components of $\vec{n}$. However, the third component, $ n_{\perp}  = \left(-\sin \chi \,\, n_x+\cos \chi \,\, n_y  \right)$, perpendicular to these two, can be indirectly inferred from the normalization condition $n_{\perp}^2 = 1 -  n_\chi^2 -n_z^2$. Since we only have access to $n_{\perp}^2$, its sign is ambiguous. Fortunately, this is irrelevant for the quantum metric computation, given that it only depends on the squared components.

Although not crucial to graphene, additional entanglement of the sublattice degree of freedom, e.g., with spin due to large spin-orbit coupling would lead to $\vec{n}^2 < 1$. As we discuss in more detail in \appref{OtherDegreesOfFreedom}, we then obtain access to a reduced form of the quantum metric.

With the third (perpendicular) component derivative obtained indirectly, we have access to the quantum metric along the scanned path,
\begin{equation}
    g_{\theta\theta} = \frac{1}{4}\left[\,\dot{n}_\chi^{2} + \dot{n}_z^{2}
+ \frac{\bigl(n_\chi\dot{n}_\chi + n_z\dot{n}_z\bigr)^{2}}
       {1 - n_\chi^{2} - n_z^{2}}\,\right], \label{gthetathetaExpr}
\end{equation}
where $\dot{n}_{\chi (z)} := \partial_\theta n_{\chi (z)} / K$. 
If either no projection to the top layer is necessary, the top-layer weight $p_S$ is approximately momentum independent (as is the case for TBG), or its momentum dependence is known, we can extract the derivatives $\dot{n}_{\chi (z)}$ as they follow from the measurement intensity derivative, $\partial_\theta \Gamma_{1,2}$, along the scanned path. We emphasize that \equsref{nChiNz}{gthetathetaExpr} are independent of $\chi$ which is thus not needed to extract $g_{\theta\theta}$ from measurements.

\subsection{Simple example}
As a concrete example illustrating the procedure, let us consider a sample described by a gapped Dirac model $h_S(\vec{k}') = v (k'_x \sigma_x + k'_y \sigma_y) + m \sigma_z - \mu_S$, similar to the tip but with $v \ll v_D$. Here, the velocity $\vec{v}_0$ that is obtained from linearizing the dispersion obeys $|\vec{v}_0| \leq v$ due to the presence of $m$. Note that the ratio between the sample and the tip's lattice defines the ratio between the length of the path probed on the sample and the twisting angle. In our example, we use a continuum model, so this ratio just defines how much of the sample's band is scanned. Effectively, on the path we probe $h_S((R_\theta-\mathbbm{1})\vec{K}_T)$ when varying $\theta$, where $R_\theta$ rotates 2D vectors by $\theta$.

\begin{figure}[tb]
    \centering
    \includegraphics[width=\linewidth]{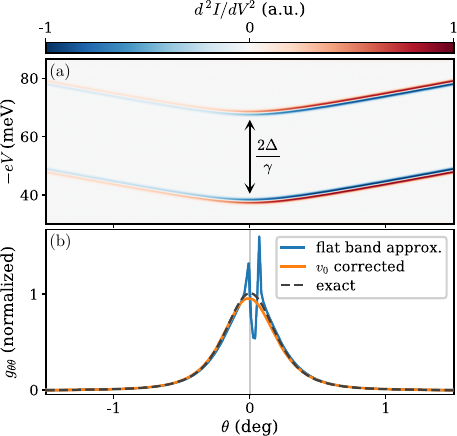}
    \caption{(a) Conductance derivative $d^2I/dV^2$ as a function of the twisting angle. (b) Exact (dashed line) and reconstructed (solid lines) quantum metric component $g_{\theta \theta}$ along the scanned line. The blue curve shows the metric reconstruction within the flat band approximation $\beta \rightarrow 0$ (for a quasi-flat band), while the orange line takes into account the finite $v_0$ effects. Here, the two roots (and therefore $g_{\theta \theta}$ solutions) coming from taking the finite $v_0$ into account are visually indistinguishable. Parameters used: $v_0/v_D = 0.045$, $m=3$ meV, $w_0/w_1=1.25$, $\Delta = 15$ meV, $\chi = \pi/4$, $\mu_S =0 $ and $\mu_T = 50$ meV.}
    \label{fig:qtm-metric}
\end{figure}

In \figref{fig:qtm-metric}, we focus only on the upper band of the sample. In the $d^2I/dV^2$ plot in \figref{fig:qtm-metric}(a), we have two features resembling the sample's band. They correspond to scanning the sample with the tip's valence and conductance bands. They are split by $2\Delta/\gamma$ as expected from the argument $\lambda[\phi_c^\lambda- \phi]= (\lambda(\xi^S_{\vec{p}_c} +eV + \mu_T) - |\Delta|/\gamma)$ of the step function in \equref{eq:nu_c}. 

To reconstruct the metric at a given $\theta$, we first fit the shape of $d^2I/dV^2$ following the distribution derived from \equref{eq:T_nu}, i.e., a Dirac delta and its derivative, which allows us to find the two critical voltages $V_c^\lambda$ that give us the resonant condition of $\phi \rightarrow \phi_c^\lambda$. Then through the fit, we extract the weight from the $\delta'(\lambda[\phi_c^\lambda- \phi])$ term; which is, in turn, proportional to the expectation value of $\hat{P_\pm}$, as defined in \equref{eq:gamma_def}. This is then used to obtain $\Gamma_{1,2}$ and perform the reconstruction protocol as introduced above. In this example, we also find that considering the curvature isotropic to obtain $\kappa$ from the extracted band is a good approximation. We note that this is particularly true near the Dirac point, where the curvature correction is most relevant. More generally, whenever we probe a $C_{3z}$ invariant point, we have $\vec{v}_0=0$ and a band extreme (if there is no crossing), where the curvature correction is relevant. Exactly at the symmetric point, the curvature is isotropic and can be extracted directly from our measurement. In the vicinity of it, $M^{-1}$ is not exactly isotropic but this is still a good approximation.

With $|T_\pm(\vec{p}_{c,n})|^2$ already determined through the first term in \equref{eq:T_nu}, fitting the second term in \equref{eq:T_nu} provides redundant information. This can be used to double-check the procedure and improve the fitting.

In \figref{fig:qtm-metric}(b), we see that the reconstructed $g_{\theta \theta}$ (solid blue line) overall has good agreement with its exact value, although we applied our protocol for perfectly flat bands ($\vec{v}_0=0$) to a system with a small but finite bandwidth ($v\neq 0$). The only exception is the region close to $\theta =0$, with significant deviations. At that point, $n_\perp $ and, thus, the denominator in the second term of \equref{gthetathetaExpr} vanish, making $g_{\theta\theta}$ much more sensitive to deviations of $n_\chi$ and $n_z$ from their exact value. This problem would arise from noisy data as well (see \appref{app:noise} for a more detailed discussion). However, we emphasize that the knowledge of the (approximate) values of $n_\chi$ and $n_z$ will allow to flag regions in experimental implementations of this procedure where errors are expected to make the extraction of $g_{\theta\theta}$ less reliable.

\subsection{Including finite dispersion}\label{FiniteDispersionSimple}
Although we cannot remediate this effect in the case of noise, we can improve our protocol by taking into account the finite bandwidth in the sample. The key consequence of $\vec{v}_0 \neq 0$ is that $\vec{k}_c^\lambda \propto |\Delta| \lambda \vec{v}_0 \neq 0$ so that the tip and, more crucially, the sample states are evaluated at slightly different momenta for $\lambda=+$ and $\lambda=-$.

While we will generalize this below, let us first assume that $\vec{v}_0$ can be approximated as parallel to the scan path since this allows taking into account the effect of the dispersion in a straightforward way. Importantly, the shift of the scanned momentum where the sample's wave function is evaluated can now be thought of as a ``mislabeling'' of the current angle: when the tip is at a given $\theta$, we are actually probing the points $\theta  + \lambda \delta \theta(k_c)$, with $\delta\theta(k_c) = k_c /K $, $k_c = |\vec{k}_c^\lambda|$. The change of the tip wave functions in \equref{tip_states_general} can be easily taken into account by modifying the projectors accordingly (e.g., $d^z_\pm  \rightarrow d^z_\pm / \gamma$, see \appref{app:expansion} for the full expressions). 

Hence, if we relabel the measured $\Gamma_\pm (\theta) \rightarrow \Gamma_\pm(\theta \mp \delta \theta)$ and use the $\vec{v}_0$-corrected projectors, we can take into account the dispersion as long as $\vec{v}_0$ is approximately aligned with the path. As the latter is true (exactly) for our minimal example studied above, we can improve the reconstruction, in particular near the pathological point with $n_\perp = 0$. We can see that through the orange curve in \figref{fig:qtm-metric}(b). We also note that the sign ambiguity in $n_\perp$ becomes a root ambiguity for finite $\vec{v}_0$. This is related to the additional term $d_{\perp} = \frac{\beta}{2}(w_0^2 - w_1^2) \sin(\chi + \varphi_0)$ in the projectors, where $\varphi_0$ is the polar angle of $\vec{v}_0$. Now with $d_{\perp}\neq0$, also $n_\perp$ enters in $\Gamma_1$ and thus ``interferes'' with $n_\chi$, which leads to two possible roots for $n_\chi$ and $n_\perp$ (see \appref{app:expansion} for details). These two roots correspond to two different Bloch vectors that move smoothly across the Bloch sphere as we twist our system. 

In some cases, the difference between these two solutions is negligible and both represent the quantum metric faithfully, as it happens in the examples that we use in this work; for instance, both roots give indistinguishable results in \figref{fig:qtm-metric}(b). The quantity that can help us identify this scenario is $d_{\perp}$. Whenever $d_\perp$ is sufficiently small (e.g., small $\beta$), both roots are again close to having the same magnitude but opposite sign and $g_{ij}$ does not depend on which root is chosen. But even if this is not applicable, our method singles out two possible $g_{ij}(\vec{k})$; to pinpoint the correct solution, one can still supplement our protocol with other methods, e.g., symmetry analysis and numerical calculations, or other measurements.

A different method to circumvent the difference of the shifts of the wavefunction for $\lambda=\pm$ is to instead fit both terms in \equref{eq:T_nu}. This will give access to $|T_\pm(\vec{p}^\lambda_{c,n})|^2$ at the \textit{same} momentum $\vec{p}^\lambda_{c,n}$. Since it involves the subleading term, this could be more prone to noise but provides a viable alternative.

\subsection{General dispersion and full tensor extraction}
Finally, we extend our protocol further which will  allow us to not only take into account $\vec{v}_0$ that are not aligned with the path but also extract the whole quantum metric and, in fact, quantum geometric tensor. To this end, we require the following. First, we need to impose some constraints on $\vec{v}_0$. The sample's band must be dispersive but still sufficiently flat such that $0<|\vec{v}_0|\ll v_D$ along the path. This will control our first-order expansion in $\beta$. In addition, we will focus on points on our path where $\vec{v}_0$ is not parallel to the path; in regions where it is (almost) aligned, we can use the approach described in \secref{FiniteDispersionSimple}. Second, we need to have access to two different values of $\Delta$.

A non-zero, non-aligned $\vec{v}_0$ implies that the conduction and valence band of the tip scan different contours; this is how we will obtain the components of $Q_{ij}$ beyond $g_{\theta\theta}$ in \equref{gthetatheta}. Due to $|\vec{v}_0|\ll v_D$, we can expand the expectation values $n_i(\vec{k}^\lambda_c)$ entering $\Gamma_{\pm}= p_S\mathcal{N}  \bra{\Psi^S (\vec{k}^\lambda_c)} \hat{P}_{\lambda} \ket{ \Psi^S(\vec{k}^\lambda_c)}$, i.e., $n_i(\vec{k}^\lambda_c) \sim n_i(0) + \lambda k_c n'_i(0)$ up to linear order in $k_c$. Here, the prime stands for the derivative $\frac{1}{|\vec{v}_0|}\vec{v}_0\cdot \nabla_{\vec{k}}$ along $\vec{v}_0$. With this, we can write the measurements as 
\begin{equation}
\label{eq:gamma_expansion}
\begin{aligned}
\Gamma_1 &\sim p_S \mathcal{N}\big[\,d_0 + d_\chi n_\chi + d_\perp n_\perp\,\big]
          \;+\; p_S\mathcal{N}\,\delta_z n_z'\;k_c, \\[10pt]
\Gamma_2 &\sim p_S \mathcal{N}\,d_z n_z
          \;+\; p_S \mathcal{N}\big[ \delta_\chi n_\chi'\,\big]\;k_c, 
\end{aligned}
\end{equation}
where all expectation values are evaluated at $\vec{k}=0$ (the scanning point of the tip). The explicit form of $d_i$ and $\delta_i$, as well as the derivation of \equref{eq:gamma_expansion} can be found in \appref{app:expansion}. 
Here is where the second requirement plays its role. The access to two $\Delta$, i.e., two values of $k_c$, allows us to distinguish the term linear in $k_c$ from the one independent of it. The linear term gives us direct access to the derivatives $n_i'$, while the other term yields the derivatives in the scan direction. Therefore, we can compute all the components $Q_{ij}$ for $i,j = \{\theta, \vec{v}_0\}$. Here, the quantum geometry components are expressed in a basis that is not necessarily orthogonal. However, we can perform a basis transformation to guarantee this property. One natural choice is to have the scan direction as an element of this new basis and transform the other components, e.g.,  $Q_{v_0 v_0} = \vec{v}_0^T Q \vec{v}_0$. The explicit expressions for each component are shown in \appref{app:othogonal}.

As before, we have to use the normalization condition to access the perpendicular component $n_\perp$. However, since $\vec{v}_0\neq 0$, we find again that in $\Gamma_{1}$, the two components $n_\chi$ and $n_\perp$ are entangled in the $k_c$ independent part, i.e., the part that gives us $\partial_\theta \vec{n}$. We thus, again find a root uncertainty.

\begin{figure}[tb]
    \centering
    \includegraphics[width=\linewidth]{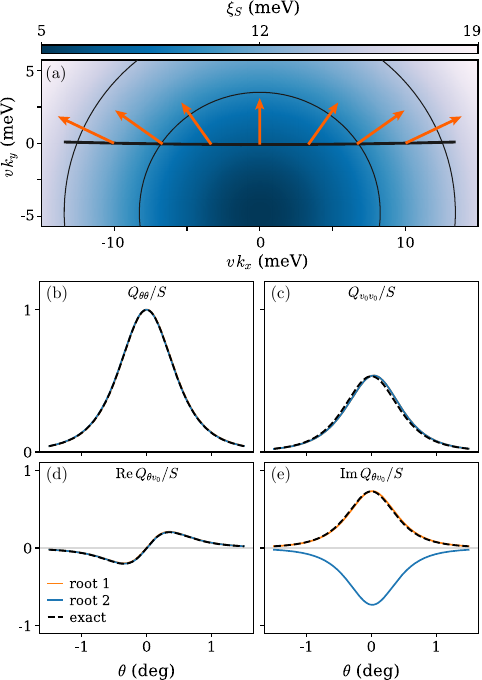}
    \caption{(a) Dispersion of the sample band we probe. The black line represents the scanned path, while the orange arrows show the local $\vec{v}_0$. Here we see that the velocity and the path are not parallel, which is a requirement to extract the whole quantum geometric tensor. (b)-(d) show the extracted components of the quantum metric and the respective exact values. For the Berry curvature, shown in (e), the two possible roots have opposite signs. The quantum geometric tensor components are normalized by the biggest value $S= \max |Q_{ij}|$. Parameters used: $\chi=0$, $\Delta = [25,35,50]$ meV, $v/v_D = 0.045 $, $k_0 = -5$ , $m=5$ meV, $\mu_T=60$meV. }
    \label{fig:3}
\end{figure}

To illustrate the full protocol with the first order expansion in $k_c$, we consider the same sample Hamiltonian as above but assume that the path is shifted by $\vec{k}_0 \neq 0$ away from its (gapped) Dirac cone, see \figref{fig:3}(a). Consequently, as opposed to before, $\vec{v}_0$ is not aligned with the path.
The magnitude of $\vec{v}_0 (\theta)$ can be extracted through $\gamma$ in the gap $2\Delta/\gamma$ separating the two scans (conductance and valence tip bands), while the direction can be extracted by analyzing the velocity component along the scan $\partial_\theta V_c^\lambda$ and comparing to the obtained magnitude. This method of extracting $\vec{v}_0$ leads to a sign indetermination. However, as above, this can be solved by supplementary information from other methods and theoretical considerations. In our example, both signs yield the same result for $g_{ij}$ but opposite signs for $\Omega = -2\text{Im} \,Q_{xy} = - \vec{n} \cdot (\partial_{k_x} \vec{n} \times \partial_{k_y} \vec{n})/2$. In this example, we use $\kappa$ from the model, since in this case the region where this correction matters the most does not correspond to a high-symmetry point. In an experimental realization, this information, as the sign of $\vec{v}_0$ can be obtained with supplementary methods. We also used three different $\Delta$ across the $25-50$ meV range.

In \figref{fig:3}(b-e), we compare the extracted quantum geometry tensor components with the exact values. We can see that for our example the two ambiguous roots are almost identical and that both of them provide good reconstructions of the metric. In the imaginary part of the quantum geometry tensor (i.e., the Berry curvature), the root ambiguity is expressed as a sign indetermination. Furthermore, we note that we avoided the pathological unbounded error in \figref{fig:qtm-metric}(b), since the scan path does not contain any point with $n_\perp \rightarrow 0$.

\section{Discussion}
In this work, we explored the potential of quantum twisting microscopy with sublattice polarized tips and showed that it provides direct access to the projectors in \equref{ProjectorsWeHaveAccessTo} measured in the sample bands along the momentum path traced out by the tip’s Dirac cones. This allows to quantify how well $C_{2z}\Theta$ and chiral symmetry $C$ are obeyed by the associated Bloch states locally in momentum space. We further showed how the quantum metric along the scanned path, see \equref{gthetatheta}, can be extracted. If two different values of the sublattice polarization $\Delta$ in the tip can be used, one can even obtain information about the momentum dependence of the states along the direction perpendicular to the path, by taking advantage of the momentum shift $k_c \neq 0$ shown in \figref{fig:1}(c). This gives access to all components of the quantum geometric tensor in \equref{QGT}. As such, our approach could provide direct experimental access to the wavefunction signatures of the complex symmetry breaking in correlated 2D systems, such as TBG. It would also fill the gap of getting access to the quantum geometry in 2D systems and would thus allow to elucidate its interplay with strong correlations. 

In practical applications of our protocol, it will be beneficial to extract $p_S\mathcal{N}$, entering \equref{nChiNz}, from measurements rather than computing it theoretically. This can be done by using consistency checks based on the reconstructed $n_z$ and $n_\chi$ along the momentum paths, supplemented by symmetry or theoretical considerations. For instance, as long as $C_{3z}$ is intact, the sublattice expectation values $(\braket{\sigma_x},\braket{\sigma_y})$ should have a vortex structure around $C_{3z}$-invariant points, such as the K points of the sample. This means that $n_\chi$ should change sign when crossing it, which constrains $p_S\mathcal{N}$ so that $n_z$ and $n_\chi$ in \equref{nChiNz} only depend on $w_0/w_1$.

Another potentially important aspect will be to verify and quantify the tip's sublattice polarized gap. This can be done by measuring the tunneling current for a simpler sample system with well-known spectra, such as single-layer graphene. To effectively obtain access to two different $\Delta$, one could use two different tips with different twist angles between graphene and hBN. Alternatively, one could take advantage of the moiré superlattice reconstruction in the tip: it gives rise to secondary \cite{PhysRevB.84.195414,jung2017moire} gaps at different energies. Besides multiple gaps, this could yield slightly different momentum paths. Finally, also AB-stacked bilayer graphene could provide a way to sublattice-polarized tips with (displacement field) tunable gaps.

Taken together, our work shows that the quantum twisting microscope is a very versatile technique that, when supplmented with polarized tips, not only gives access to electronic bands but also to their associated momentum-dependent wavefunctions.

\begin{acknowledgments}
M.S.S.~thanks D.~Efetov and D.~Sedov for discussions on QTM. Both authors further acknowledge funding by the European Union (ERC-2021-STG, Project 101040651---SuperCorr). Views and opinions expressed are however those of the authors only and do not necessarily reflect those of the European Union or the European Research Council Executive Agency. Neither the European Union nor the granting authority can be held responsible for them.
\end{acknowledgments}

\bibliography{draft_Refs}

\onecolumngrid

 \begin{appendix}

\section{Contour calculation}
\label{app:contour}

We start from the equation defining the energy contour in the vicinity of a given Dirac point,
\begin{equation}
\label{eq:contour}
   \xi^S_0 + \vec{v}_0 \cdot \vec{k} + \frac{1}{2}\vec{k} \cdot M^{-1} \vec{k} + \mu_S  -\lambda \sqrt{( v_D \vec{k})^2+ \Delta^2} = \phi.
    \end{equation}
For convenience, let us define $\vec{v}_0 = v_0\, \hat{\vec{x}}$, $A=\xi^S_0 + \mu_S - \phi$ and square the equation that defines the integration contour (i.e., intersection between bands)

\begin{equation}
\begin{split}
   \left(A+v_0 k_x\right)^2-\left(v_D k\right)^2-\Delta^2+\left(A+v_0 k_x\right) \vec{k} M^{-1} \vec{k}+\frac{1}{4}\left(\vec{k} M^{-1} \vec{k}\right)^2=0
   \end{split}
    \end{equation}
note that we must have $\lambda (A + \vec{v}_0 \cdot \vec{k})>0$. This condition is not explicit in the squared equation but it is important to reintegrate it to the contour expression in the end. We keep terms up to quadratic order in $\vec{k}$ and obtain
\begin{equation}
    \left(A+\vec{v}_0 \cdot \vec{k}\right)^2-\left(v_D k\right)^2-\Delta^2+A \vec{k} M^{-1} \vec{k}=A^2-\Delta^2+2 A \vec{v}_0 \cdot \vec{k}-\vec{k} B_A \vec{k}=0,
\end{equation}
with $B_A \equiv Q - AM^{-1}$ and  $Q= v_D^2 \mathbbm{1} - \vec{v_0}\vec{v_0}^T$.

Now, we recast the equation above as an ellipse equation:
\begin{equation}
\left(\vec{k} - \vec{k}_c(A)\right) B_A \left(\vec{k} - \vec{k}_c(A)\right) = R^2(A)
\end{equation}
with the center of the ellipse being $\vec{k}_c(A) = A B_A^{-1} \vec{v}_0$ and the radius $R^2(A) = A^2\left(1 + \vec{v}_0 B_A^{-1} \vec{v}_0\right) - \Delta^2$. We can further rewrite these expressions as
\begin{equation}
    \begin{split}
        \vec{k}_c(A) &= \left(\mathbf{1} + \frac{A}{v_D^2} M^{-1}\right)\vec{k}_{c,0}(A),\\
        R^2(A) &= A^2\gamma^2 - \Delta^2 + A\,\vec{k}_c^0(A) M^{-1} \vec{k}_c^0(A),
    \end{split}
\end{equation}
up to first order in $M^{-1}$, with $\gamma = (1 - v_0^2/v_D^2)^{-1/2}$. Here, $\vec{k}_{c,0}(A) \equiv A\gamma^2 \vec{v}_0 / v_D^2$ is the center of the ellipse that one would obtain without the sample curvature, i.e., for $M^{-1}=0$. Note that $\vec{k}_{c,0}(A)$ can be $\pm |A|\gamma^2 \vec{v}_0 / v_D^2$ depending on the condition $\lambda (A + \vec{v}_0 \cdot \vec{k})>0$, while the curvature correction is the equal for both $\lambda$. For a given sample band, the center of the ellipse will then be defined by whether the tip is tuned to the conductance or valence band. In the limit of the contour shrinking to a point, we have $\vec{k} \rightarrow \vec{k}_c$: $\lambda (A + \vec{v}_0 \cdot \vec{k}_c)\geq0$, which gives us $\lambda (A + A(\gamma^2-1 ))\geq0$ and the condition $\sign
\,A = \lambda$.

The vanishing condition, i.e., the condition $A= A_c$ such as the radius becomes zero $R(A_c) = 0 $, can be written as $A_c = A_{c,0} + \delta A_c$ where the first part corresponds to the solution for the case without the curvature correction, i.e., $M^{-1}=0$. Then one can obtain $A_{c,0} = \lambda \frac{|\Delta|}{\gamma}$, $\vec{k}_c^\lambda \equiv \vec{k}_{c,0}(A_{c,0}) = \frac{\lambda \Delta \gamma}{v_D^2}\vec{v}_0$ and the correction $\delta A_c = -\frac{\vec{k}_c^\lambda M^{-1} \vec{k}_c^\lambda}{2\gamma^2}$. Hence, we have
\begin{equation}
    A_c = \lambda \frac{|\Delta|}{\gamma} - \frac{1}{2}\vec{k}_c^\lambda M^{-1}\vec{k}_c^\lambda,
\end{equation}
and the energy condition
\begin{equation}
    \xi^S_0 + \mu_S - \frac{\lambda|\Delta|}{\gamma} +\frac{\vec{k}_c^\lambda M^{-1} \vec{k}_c^\lambda}{2\gamma^2}   = \phi_c^\lambda.
\end{equation}

For our goals, the three important quantities derived here are: the center of the ellipse, i.e., the point $\vec{k_c}(A)$ that the contour shrinks to; the shape of the contour, described by $B_A$; and the critical energy $\phi_c^\lambda$. From our calculations, we can see that they can be expressed as a sum of a $m_T M^{-1}$ independent and dependent (curvature correction) term. We can also observe that the only curvature correction that survives $\vec{v}_0 \rightarrow 0$ is the one in the shape of the contour. For the purposes of this work, it suffices to only keep the correction in $B_A$, which is equivalent of neglecting terms of order $O(m_T M^{-1} \beta )$ and $O(m_T M^{-1} \beta^2 )$. 

These considerations lead us to
\begin{equation}
    \begin{split}
             \phi_c^\lambda= \xi^S_{0} + \mu_S -\frac{\lambda|\Delta|}{\gamma}, \,\text{ and }\,  \vec{k}^\lambda_c = \frac{\lambda |\Delta| \gamma} {v_D^2} \vec{v}_0,
    \end{split}
\end{equation}
as stated in \equref{CricialValueofPhi} of the main text, while the shape of the contour will be used in the following section.

\section{Contour density of states}
\label{app:JDOS}

To compute $\nu_c^\lambda$ from \equref{eq:JDOS}, we can first write it as
\begin{equation}
     \nu^\lambda_c=\int \frac{d^2 p}{(2\pi)^2}\delta(\lambda [\delta\epsilon_p^{ \lambda}-\phi]) 
\end{equation}
and then use the fact that
\begin{equation}
    -\lambda \delta(\lambda [\delta\epsilon_p^{ \lambda}-\phi]) =  \frac{\partial}{\partial \phi} \Theta(\lambda [\delta\epsilon_p^{ \lambda}-\phi]),
\end{equation}
to write the expression
\begin{equation}
    \nu_c^\lambda = - \frac{\lambda}{(2\pi)^2}\frac{\partial}{\partial \phi} \int d^2 p\,\,\Theta(\lambda [\delta\epsilon_p^{ \lambda}-\phi]).
\end{equation}
For a single contour, the integral on the RHS is just the area of the ellipse defined by $\phi$, see \appref{app:contour}, $\pi R^2(A)/\sqrt{\det B_A}$ with $R^2(A)=A^2\gamma^2-\Delta^2$ and $B_A = Q - A M^{-1}$. To first order in $M^{-1}$, $\det B_A = \det Q\,[1 - A\,\mathrm{tr}(Q^{-1}M^{-1})]$ with $\det Q = v_D^4/\gamma^2$, and
\begin{equation}
    Q^{-1} = \frac{1}{v_D^2}\big(\mathbf{1} + \beta^2\gamma^2\,\hat{\vec v}_0\hat{\vec v}_0^T\big)
    \quad\Rightarrow\quad
        \mathrm{tr}(Q^{-1}M^{-1}) = \frac{1}{|\Delta|\gamma}\Big[\mathrm{tr}\big(m_T M^{-1}\big)
    + \beta^2\gamma^2\,\hat{\vec v}_0\, m_T M^{-1}\hat{\vec v}_0\Big].
\end{equation}
The second term is a $\beta^2$ correction to a term that is already first order in the curvature, i.e.  of order $m_T M^{-1} \beta^2$, the same order of the corrections neglected in \appref{app:contour} we drop it as well. Hence
\begin{equation}
    \int d^2 k\,\,\Theta(\lambda [\delta\epsilon_k^{ \lambda}-\phi])
    = \frac{\pi \gamma (A^2 \gamma^2 - \Delta^2)}{ v_D^2}
      \left(1 + \frac{A\,\mathrm{tr}M^{-1}}{2 v_D^2}\right)
    = \frac{\pi \gamma (A^2 \gamma^2 - \Delta^2)}{ v_D^2}
      \left(1 + \kappa\,\frac{\gamma A}{|\Delta|}\right),
\end{equation}
where in the last step we inserted $\mathrm{tr}M^{-1} = 2\kappa/m_T$ with $m_T = |\Delta|\gamma/v_D^2$. Now, from $A=\xi^S_0 + \mu_S - \phi$, we have $dA/d\phi = -1$ and thus
\begin{equation}
\label{eq:nu_of_A}
    \nu^\lambda_c = \frac{|A|\,\gamma^3}{2\pi v_D^2}
    \left[1 + \lambda\kappa\,\frac{3\gamma^2 A^2 - \Delta^2}{2\gamma |\Delta|\, |A|}\right]
    \Theta(|A|- |\Delta|/\gamma).
\end{equation}
Note that if the contour (intersection) does not exist, the area is zero and so is the derivative. We account for that through the extra factor $\Theta(|A|- |\Delta|/\gamma)$. In contrast to the gapless tip considered in previous works \cite{inbar2023quantum,wei2025dirac}, our $\nu_c^\lambda$ is discontinuous at the point where the elliptical contour shrinks to a point. We can further work on the argument of the $\Theta$ function by rewriting it in
terms of $\phi$ and $\phi_c^\lambda$. From
\begin{equation}
    \xi^S_0 + \mu_S   = \phi_c^\lambda+\lambda\frac{|\Delta| }{\gamma},
\end{equation}
and
\begin{equation}
    A=\xi^S_0 + \mu_S - \phi,
\end{equation}
we get
\begin{equation}
\begin{split}
A = \phi_c^\lambda +\lambda\frac{|\Delta| }{\gamma} - \phi\\
\lambda A = \lambda(\phi_c^\lambda- \phi) + \frac{|\Delta|}{\gamma}\\
\lambda A  - \frac{|\Delta|}{\gamma} = \lambda(\phi_c^\lambda- \phi).
\end{split}
\end{equation}
We also recall (\appref{app:contour}) that $|A| = \lambda A$, so that
\begin{equation}
    |A| = \frac{|\Delta|}{\gamma} + |\phi_c^\lambda - \phi| .
\end{equation}
Inserting this into \equref{eq:nu_of_A}, the prefactor and the bracket combine into
\begin{equation}
    |A|\left[1 + \lambda\kappa\,\frac{3\gamma^2 A^2 - \Delta^2}{2\gamma |\Delta|\, |A|}\right]
    = |A| + \lambda\kappa\left(\frac{3\gamma |A|^2}{2|\Delta|} - \frac{|\Delta|}{2\gamma}\right)
    = \frac{|\Delta|}{\gamma}(1+\lambda\kappa) + (1+3\lambda\kappa)\,|\phi_c^\lambda - \phi|
      + \frac{3\lambda\kappa\gamma}{2|\Delta|}\,|\phi_c^\lambda - \phi|^2 ,
\end{equation}
so that
\begin{equation}
\nu^\lambda_c = \frac{ \gamma^2 |\Delta|\,(1+\lambda\kappa)
                     + \gamma^3|\phi_c^\lambda - \phi|\,(1+3\lambda\kappa)
                     + \tfrac{3}{2}\lambda\kappa\gamma^4\,|\phi_c^\lambda - \phi|^2/|\Delta|}{2 \pi v_D^2}\,
               \Theta(\lambda [\phi_c^\lambda-\phi]) .
\end{equation}
The curvature of the sample band enters the discontinuity through the factor $1+\lambda\kappa$, which counts the additional (missing) states per energy of the relative dispersion $\delta\epsilon^\lambda_k$ when the sample band curves in the same (opposite) sense as the tip band $\lambda$. Furthermore, for $|\phi - \phi^\lambda_c| \ll |\Delta|$, i.e. $\phi \rightarrow \phi_c^\lambda$, we can discard the last part and write
\begin{equation}
\nu^\lambda_c = \frac{ \gamma^2 |\Delta|\,(1+\lambda\kappa)
                     + \gamma^3|\phi_c^\lambda - \phi|\,(1+3\lambda\kappa)
                     }{2 \pi v_D^2}\,
               \Theta(\lambda [\phi_c^\lambda-\phi])
\end{equation}
as in the main text.

\section{Effect of noise}
\label{app:noise}
Let us analyze the effects of measurement error, i.e., noise in the current $I(V)$ extraction, in our protocol. We start by looking at the derivative of the normalization condition $n_\chi^2 + n_\perp^2 +n_z^2 = 1$ along the scanned direction,
\begin{equation}
    n_\chi \dot{n}_\chi +n_z \dot{n}_z + n_\perp \dot{n}_\perp = 0,
\end{equation}
which gives us $\dot{n}_\perp = \frac{-n_\chi \dot{n}_\chi +n_z \dot{n}_z}{n_\perp}$. Note that for the true $n_i$ components, which obey the normalization relation, the limit $n_\perp  \rightarrow 0$ yields a removable singularity in $\dot{n}_\perp$ and it remains finite. However, once the measurements of $n_\chi,n_z$ introduce errors $\epsilon_{\chi(z)}$, we propagate it to $n_\perp$
\begin{equation}
    \delta\dot n_\perp
    =-\frac{1}{n_\perp}\left(\dot n_\chi\epsilon_\chi+\dot n_z\epsilon_z
    +n_\chi\dot\epsilon_\chi+n_z\dot\epsilon_z\right)
    +\frac{\dot n_\perp}{n_\perp^{2}}
    \left(n_\chi\epsilon_\chi+n_z\epsilon_z\right),
\end{equation}
and the quantum metric
\begin{equation}
    g_{\theta\theta} = \frac{1}{4}\left[\,\dot{n}_\chi^{2} + \dot{n}_{\perp}^{2}
+ \dot{n}_z^{2}\right] +\delta_g^b + \delta_g^u.
\end{equation}
Here, we split the error in $g_{\theta \theta}$ into a bounded $\delta_g^b$ and unbounded  $\delta_g^u$ part
\begin{equation}
    \delta_g^{b}
    =\frac{1}{2}\left(\dot n_\chi\dot\epsilon_\chi+\dot n_z\dot\epsilon_z\right),
\end{equation}
\begin{equation}
    \delta_g^{u}
    =-\frac{\dot n_\perp}{2n_\perp}
    \left(\dot n_\chi\epsilon_\chi+\dot n_z\epsilon_z
    +n_\chi\dot\epsilon_\chi+n_z\dot\epsilon_z\right)
    +\frac{\dot n_\perp^{2}}{2n_\perp^{2}}
    \left(n_\chi\epsilon_\chi+n_z\epsilon_z\right).
\end{equation}
While $\dot n_\perp$ remains finite always, once we scan a point $\theta$ such as $n_\perp\rightarrow0$, the unbounded part $\delta^u_g$ dominates and corrupts the signal. Note that this condition is directly linked to the parameter $\chi$ in the tunneling matrix in \equref{eq:tunn_matrix}, which defines $ n_{\perp}  = \left(-\sin \chi \,\, n_x+\cos \chi \,\, n_y  \right)$.

\section{Expansion for dispersive bands}
\label{app:expansion}
\subsection{General procedure}\label{GeneralProcedureInAppendix}
Let us develop our protocol for a given critical momentum point $\vec{k}^\lambda_{c}$ considering a first order expansion in $k_c = |\vec{k}^\lambda_{c}|$ (the same steps follow for the generic $\vec{p}_{c,n}$). We start from the approximation for the current
\begin{equation}
I\approx 2 \pi e \Omega  \sum_{n,\lambda} \left[f\left(\xi_{\vec{k}^\lambda_{c,n} }^S\right)-f\left(\xi_{\vec{k}^\lambda_{c,n} \lambda}^T\right)\right] \left\langle\left|T_{\lambda,n}\right|^2 \right\rangle_{\phi}\,\, \nu^\lambda_{c,n}(\phi),
\end{equation}
we only need to study the expansion of $\left\langle\left|T_{\lambda,n}\right|^2 \right\rangle_{\phi}\,\, \nu^\lambda_{c,n}(\phi)$ for a given contour $\phi \rightarrow \phi_c^\lambda$ (given that all $\vec{p}_{c,n}$ are equivalent, due to symmetry, as in the main text). In principle, one should also use the expanded factor $f(\xi_{\vec{k}^\lambda_{c,n} }^S)-f(\xi_{\vec{k}^\lambda_{c,n} \lambda}^T) \approx  f(\xi^S_0) - f(\xi^T_{0,\lambda}) + \lambda v_0 k^\lambda_{c} \, f'(\xi^S_0)$. However, we can always choose $\mu_S$ and $\mu_T$ such as: (i) $f'(\xi^S_0)=0$, which happens when the sample band of interest is either completely full or empty and (ii) both $\lambda$ give the same $f(\xi^T_{0,\lambda})$, which happens when the tip's Fermi level is far away from the band edges, i.e., $\mu_T - |\Delta| \gamma \gg k_B T$. For the scenario with the sample filled (empty), we  would need the tip to be mostly empty (full) to get a finite current.

Before we work on the $\vec{k}_c$ expansion, note that the tip states involved in the tunneling are not the fully polarized states, rather, we have
\begin{equation}
\label{tip_states_general}
    \zeta_{\lambda}(\vec{k}) = \frac{1}{\sqrt{2}} \begin{pmatrix} \sqrt{ 1+ \lambda \cos \alpha_k} \\
    \lambda e^{i \varphi_{\vec{k}}} \sqrt{1 - \lambda \cos \alpha_k}
    \end{pmatrix}.
\end{equation}
with $\tan\alpha_{k_c} = \frac{ v_D\, k_c}{\Delta} = \frac{\gamma v_0}{v_D}$, for $\Delta>0$. Furthermore, $\vec{v}_0$, points in an arbitrary direction, which defines the polar angle $\varphi_0$.  For a given contour specified by $\phi$ and some fixed $n$ (which we suppress in the following equations), this means that
\begin{equation}
    \big\langle |T_{\lambda}|^2\big\rangle_\phi
  = \Theta\big(\lambda(\phi_{c}^\lambda-\phi)\big) \left[ \frac{1+\cos\rho}{2}\,|T_\lambda(\vec{k}^\lambda_{c})|^2
  + \frac{1-\cos\rho}{2}\,|T_{-\lambda}(\vec{k}^\lambda_{c})|^2 \right],
\end{equation}
with $\cos\rho = \frac{|\Delta|}{|\Delta|+\lambda(\phi^\lambda_c-\phi)}$. Here both $\lambda$ show up for a given contour because the tip state deviates from its form at $\vec{k}_c^\lambda$ when $\lambda(\phi_{c}^\lambda-\phi)>0$ and $\big\langle |T_{\lambda}|^2\big\rangle_\phi$ can be expressed as a combination of both polarizations. Inserting $\cos\rho$,
\begin{equation}
  \big\langle |T_{\lambda}|^2\big\rangle_\phi
  = \Theta\big(\lambda(\phi_{c}^\lambda-\phi)\big) \left[ |T_\lambda(\vec{k}^\lambda_c)|^2
  - \frac{|\phi^\lambda_c-\phi|}{2(|\Delta|+|\phi^\lambda_c-\phi|)}\Big(|T_\lambda(\vec{k}^\lambda_c)|^2 - |T_{-\lambda}(\vec{k}^\lambda_c)|^2\Big)\right] .
\end{equation}
Note that this factor depends on $(\phi-\phi^\lambda_c)$; while we are interested in the limit $\phi \rightarrow \phi_c$, we cannot focus on $\big\langle |T_\lambda(\vec{k})|^2\big\rangle_{\phi_c^\lambda}$ if we want to keep terms up to linear order in $(\phi-\phi^\lambda_c)$. We rather have to expand the whole factor $\left\langle\left|T_{\lambda}\right|^2 \right\rangle_{\phi}\,\, \nu^\lambda_{c,n}(\phi)$. We then obtain
\begin{equation}
\left\langle\left|T_{\lambda,n}\right|^2 \right\rangle_{\phi}\, \nu^\lambda_{c,n}(\phi) = \frac{\Theta\big(\lambda(\phi_{c}^\lambda-\phi)\big)}{2\pi v_D^2} \gamma^2 \Big[
|\Delta|(1+\lambda\kappa)\,|T_\lambda(\vec{k}^\lambda_c)|^2
+\gamma |\phi_{c,n}-\phi|\left((1+3\lambda\kappa)\,|\bar T_{\lambda}|^2
+ \lambda\kappa\,|\delta T_\lambda|^2\right)
\Big]
\end{equation}
with $ |\overline{T}_{\lambda}|^2 = (|T_+(\vec{k}^\lambda_{c})|^2+|T_-(\vec{k}^\lambda_{c})|^2)/2$ and $|\delta T_\lambda|^2=(|T_+(\vec{k}^\lambda_{c})|^2-|T_-(\vec{k}^\lambda_{c})|^2)/2$. 
As mentioned in the text, we are interested in extracting $|T_\lambda(\vec{k}^\lambda_c)|^2$, so we analyze the term proportional to $\Delta$, which is proportional to the $\Gamma_\lambda$ defined in the main text. We regain the expression in \equref{eq:T_nu} by taking $\gamma \approx 1 + O(k_c^2)$.

Now, we finally do the expansion of $|T_{\lambda,n}|^2$ in $k_c^\lambda$. We can use that $ \Gamma_{\lambda} = \mathcal{N}  \bra{\Psi^S} \hat{P}_{\lambda} \ket{ \Psi^S}$ as in the main text [\equref{eq:gamma_def}] with $\hat{P}_\pm = d_0\,\sigma_0 + \vec{d}_\pm\cdot\vec{\sigma}$ and
\begin{equation}
\begin{aligned}
d^0 &= \tfrac12\bigl(w_0^2+w_1^2\bigr)
      + w_0 w_1\,\beta\cos(\chi+\varphi_0),\\[2pt]
d^x &= w_0 w_1\cos\chi
      + \tfrac{\beta}{2}\bigl[\,w_0^2\cos(2\chi+\varphi_0)+w_1^2\cos\varphi_0\,\bigr],\\[2pt]
d^y &= w_0 w_1\sin\chi
      + \tfrac{\beta}{2}\bigl[\,w_0^2\sin(2\chi+\varphi_0)-w_1^2\sin\varphi_0\,\bigr],\\[2pt]
d^z_\pm &= \pm\,\frac{w_0^2-w_1^2}{2\gamma}.
\end{aligned}    
\end{equation}
To overcome the fact that the two measurements take into account different sample states, i.e., $\Gamma_{\pm} = \mathcal{N}  \bra{\Psi^S (\vec{k}^\pm _c )} \hat{P}_{\pm} \ket{ \Psi^S (\vec{k}^\pm _c )}$, we use the first order expansion in $k_c$ to write
\begin{equation}
\begin{split}
    n_i(\vec{k}^\pm _c) \equiv  \bra{\Psi^S (\vec{k}^\pm _c )} \sigma_i \ket{ \Psi^S (\vec{k}^\pm_c )} &\approx n_i(0) \pm k_c \,([\hat{\vec{v}}_0\cdot \nabla] n_i )\bigg|_{\vec{k}=0}\\
    &= n_i(0) \pm k_c \,n_i'(0).
    \end{split}
\end{equation}
Here we defined $\hat{\vec{v}}_0 = \vec{v}_0/|\vec{v}_0|$ and $f'(0) = \,([\hat{\vec{v}}_0\cdot \nabla] f(\vec{k)} )\big|_{\vec{k}=0}$. Note that this expansion requires that $n_i$ are smooth in the relevant region. Now, we plug in our approximation into the linear combinations $\Gamma_{1,2} = \frac{1}{2}(\Gamma_+ \pm \Gamma_-)$ and only keep terms up to linear order in $k_c$
\begin{equation}
\begin{aligned}
\Gamma_1 / \mathcal{N} &= d_0\, + d_\chi\,n_\chi + d_\perp\,n_\perp
            +k_c\,\, \delta_z n_z', \\[6pt]
\Gamma_2 /\mathcal{N} &= d_z\,n_z
            + k_c \,\,\delta_\chi \,n_\chi',
\end{aligned}
\end{equation}
with
\begin{equation}
    \begin{aligned}
d_0 &= \tfrac12(w_0^2+w_1^2) + \beta\,w_0 w_1\cos(\chi+\varphi_0),\\
d_\chi &= w_0 w_1 + \tfrac{\beta}{2}(w_0^2+w_1^2)\cos(\chi+\varphi_0),\\
d_\perp &= \tfrac{\beta}{2}(w_0^2-w_1^2)\sin(\chi+\varphi_0),\\
d_z &= \,\frac{w_0^2-w_1^2}{2\gamma},
\end{aligned}
\end{equation}
and the 0th order expansion of $\vec{d}$ for the terms that are already proportional to $k_c$ (due to the linear expansion of $n_i(\vec{k}^\lambda_c)$)
\begin{equation}
    \begin{aligned}
\delta_\chi = w_0 w_1,\qquad
\delta_\perp = 0,\qquad
\delta_z = \frac{w_0^2-w_1^2}{2}.
\end{aligned}
\end{equation}

As stated in the main text, if one can access two values of $k_c$, i.e., two $\Delta$ values, we can isolate the linear part from the other, i.e., we have access to $A_{1,2}$ and $B_{1,2}$ separately, as defined below
\begin{equation}
\begin{aligned}
\Gamma_1 &= \underbrace{\mathcal{N}\big[\,d^0+ d_\chi n_\chi + d_\perp n_\perp\,\big]}_{\displaystyle A_1}
          \;+\; \underbrace{\mathcal{N}\,\delta^z n_z'}_{\displaystyle B_1}\;k_c, \\[10pt]
\Gamma_2 &= \underbrace{\mathcal{N}\,d^z n_z}_{\displaystyle A_2}
          \;+\; \underbrace{\mathcal{N}\big[ \delta_\chi n_\chi'\,\big]}_{\displaystyle B_2}\;k_c.
\end{aligned}
\end{equation}
Note that $B_1$ gives us direct access to $n_z'$, i.e., the derivative along $\vec{v}_0$, $A_2$ gives us $n_z$ (and hence the derivative along the scanned path $\dot n_z$ once we scan the sample) and $B_2$ gives us $n_\chi'$. Therefore, we can obtain $n_\perp'$ through the normalization condition.

However, we still need to disentangle $n_{\perp}$ from $n_{\chi}$ (and with that their $\theta$ derivatives) in $A_1$. The problem is that using the normalization condition gives us two possible pairs of $(n_{\chi}, n_{\perp})$, in contrast to the flat band limit, where the ambiguity was only the sign of $n_\perp$, which was irrelevant for the quantum metric component $g_{\theta \theta}$. These two possibilities correspond to the upper and lower sign in the following expressions:
\begin{equation}
\label{eq:two_ns}
\begin{aligned}
n_\chi^{(1,2)}  &= \frac{\left(\dfrac{A_1}{\mathcal{N}} - d^0\right) d_\chi
   \;\mp\; d_\perp\sqrt{\,(d_\chi^2 + d_\perp^2)(1 - n_z^2) - \left(\dfrac{A_1}{\mathcal{N}} - d^0\right)^{2}}}
   {d_\chi^2 + d_\perp^2}, \\[12pt]
n_\perp^{(1,2)} &= \frac{\left(\dfrac{A_1}{\mathcal{N}} - d^0\right) d_\perp
   \;\pm\; d_\chi\sqrt{\,(d_\chi^2 + d_\perp^2)(1 - n_z^2) - \left(\dfrac{A_1}{\mathcal{N}} - d^0\right)^{2}}}
   {d_\chi^2 + d_\perp^2}.
\end{aligned}
\end{equation}

As a sanity check, note that if we take the flat band limit $v_0/v_D \rightarrow 0$, we have $d_\perp =0$ and $A_1 = \mathcal{N} [d^0 + d_\chi n_\chi]$, so that \equref{eq:two_ns} reduces to $n_\chi = n_\chi$ and $n_\perp = \pm  \sqrt{ 1 - n_z^2 - n_\chi^2}$, in line with the protocol for flat bands detailed in the main text.

\subsection{Explicit plot of the Bloch components for our model}
\begin{figure}[h]
    \centering
    \includegraphics[width=\linewidth]{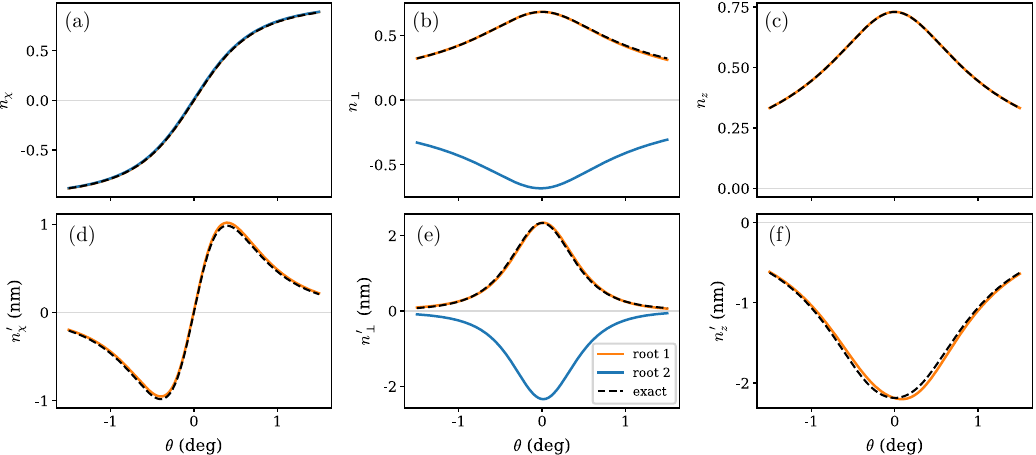}
    \caption{(a)-(c) Bloch components as a function of $\theta$ for the same model used in \figref{fig:3}. (d)-(f) are the derivatives of the Bloch components in the direction of $\vec{v}_0$. Note that $n_\chi$ and $n_\perp$ in (a) and (b) have a root ambiguity, while $n'_\perp$ in (e) only has a sign ambiguity.}
    \label{fig:app_bloch}
\end{figure}

To illustrate the two roots in more detail, we show in \figref{fig:app_bloch} the exact Bloch components and those extracted from our protocol used in \figref{fig:3} of the main text. We note that although $n_\chi$ and $n_\perp$ have a root ambiguity, due to the parameters used, both solutions are almost identical for $n_\chi$ and almost opposite for $n_\perp$. 

\subsection{Quantum geometry in an orthogonal basis}\label{app:othogonal}
Now, let us see how one can go from the basis defined and parametrized through the scan and velocity $(\theta, \vec{v}_0)$ to an orthogonal basis. 
For a small sample Brillouin zone, this direction is approximately fixed, otherwise, our transformation defines a local basis.

First, let us define $\hat{\vec{t}}$ as the tangent of the scan path, such as $\partial_\theta = |\vec{k}| \hat{\vec{t}} \cdot \nabla_{\vec{k}}$. We can write $\partial_\theta \vec{n} = |\vec{k}|(t_x \partial_x \vec{n} + t_y \partial_y \vec{n} )$ and $\vec{n}'\equiv \frac{\partial_{\vec{v}_0}}{|\vec{v}_0|} \vec{n} = \frac{v_{0x} \partial_x \vec{n} + v_{0y} \partial_y \vec{n}}{|\vec{v}_0|} $. Hence, we have
\begin{equation}
\begin{split}
    \begin{pmatrix}
        \partial_\theta \vec{n}\\
    \partial_{\vec{v}_0} \vec{n}
    \end{pmatrix} &= A
    \begin{pmatrix}
        \partial_x \vec{n}\\
    \partial_y \vec{n}
    \end{pmatrix},
\end{split}
\end{equation}
with
\begin{equation}
    A= \begin{pmatrix} |\vec{k}|\,\hat{\mathbf{t}}^{\mathsf{T}} \\[2pt]
                      \hat{\mathbf{v}}_{0}^{\mathsf{T}} \end{pmatrix}.
\end{equation}
Now, applying this transformation for the quantum metric tensor, we have $Q = A^{-1} \tilde{Q} (A^{-1})^T$, where $Q$ is expressed in the $(\hat{\vec{x}},\hat{\vec{y}})$ basis and $\tilde{Q}$ is expressed in the $(\hat{\vec{t}},\vec{v}_0)$ basis. For concreteness, we choose the path along the $k_x$ direction, i.e., changing $\theta$ corresponds to varying $k_x$. Now the quantum metric reads
\begin{align}
  Q_{xx} &= \frac{Q_{\theta\theta}}{\vec{k}^2}, \\
  Q_{xy} &= -\cot\psi\, \frac{Q_{\theta\theta}}{\vec{k}^2} + \frac{Q_{\theta v_0}}{|\vec{k}|\sin\psi}, \\
  Q_{yy} &= \cot^{2}\!\psi\, \frac{Q_{\theta\theta}}{\vec{k}^2}
            - \frac{\left(Q_{\theta v_0}+Q_{v_0\theta}\right)\cot\psi}{|\vec{k}|\sin\psi}
            + \frac{Q_{v_0v_0}}{\sin^{2}\!\psi}.
\end{align}
with $\psi$ being the angle between $\vec{v}_0$ and $\vec{\hat{t}}$. Note that if $\vec{v}_0 \parallel \hat{\vec{t}}$, $\psi =0$ and this transformation is ill defined, as expected.

\begin{figure}
    \centering
    \includegraphics[width=0.6\linewidth]{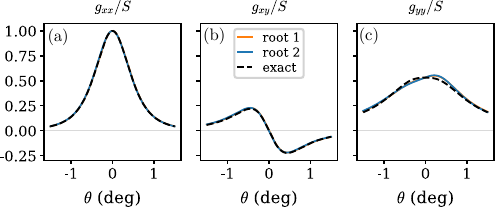}
    \caption{(a)-(c) Quantum metric components in an orthogonal base. The values components are normalized by the biggest value $S= \max g_{ij}$. All the parameters are the same as in \figref{fig:3}.}
    \label{fig:placeholder}
\end{figure}

\section{Consequences of projection and of sensitivity to sublattice degrees of freedom}\label{ProjectionAndSublatticeDOFs}

As already discussed in the main text, electrons tunnel primarily into the top layer of the sample due to the exponential decay of the wave functions such that $\ket{\Psi^S_{\vec{k}}}$ should be thought of as the projection (and subsequent renormalization) of the Bloch band of the sample to the top layer. Another important aspect is that the tunneling operator $\hat{T}$ is non-trivial in sublattice space but trivial in the spin index. In this appendix, we will discuss in detail how these two aspects---projection and insensitivity to degrees of freedom beyond sublattice---affect what exact form of quantum metric our approach provides access to for a general sample Bloch band.

To be precise, denoting the (normalized) sample Bloch band by $\ket{\Psi_{\vec{k}}}$, we are ultimately interested in the quantum metric defined as 
\begin{equation}
    g_{ij} := \text{Re} \left[ \bra{\partial_{k_i} \Psi} (1 - \ket{\Psi}\bra{\Psi}) \ket{\partial_{k_j} \Psi} \right] = \text{Re} \left[ \text{tr} \left[ P \partial_{k_i} P \partial_{k_j} P \right] \right], \quad P = \ket{\Psi}\bra{\Psi}. \label{FullQuantumMetric}
\end{equation}
For notational simplicity, we here suppressed the momentum argument $\vec{k}$ of the wavefunctions and the projectors. Note that $g_{ij}$ is U(1) gauge invariant, i.e., does not change under 
\begin{equation}
    \ket{\Psi_{\vec{k}}}\,\, \longrightarrow \,\, e^{i\alpha_{\vec{k}}}\ket{\Psi_{\vec{k}}}, \quad \alpha_{\vec{k}} \in \mathbb{R}, \label{GaugeTransformationBlochStates}
\end{equation}
which is most directly seen in the second form in \equref{FullQuantumMetric} as it involves the gauge-invariant projectors.

\subsection{Projection}
To first focus on the effect of projection onto a subspace, we split the Bloch state according to
\begin{equation}
    \ket{\Psi_{\vec{k}}} = \mathcal{P} \ket{\Psi_{\vec{k}}} + (1-\mathcal{P})\ket{\Psi_{\vec{k}}} \equiv \sqrt{p_S(\vec{k})} \ket{\Psi^S_{\vec{k}}} + \sqrt{p_{\bar{S}}(\vec{k})} \ket{\Psi^{\bar{S}}_{\vec{k}}}, \label{SplittinUpTheWavefunction}
\end{equation}
where $\mathcal{P}$ a projection operator onto the above-mentioned subspace. In our case here, this would be the top layer of the sample and, in the case of TBG, the projection on the top layer and respective momentum-lattice point $\vec{g}=\vec{g}_0$; but we keep our discussion general for now. We will only use that $\mathcal{P}$ [unlike $P$ in \equref{FullQuantumMetric}] does not depend on momentum. In \equref{SplittinUpTheWavefunction}, we also already introduced the renormalized projected Bloch bands
\begin{align}\begin{split}
    \ket{\Psi^S_{\vec{k}}} &= \frac{1}{\sqrt{p_S(\vec{k})}} \mathcal{P} \ket{\Psi_{\vec{k}}}, \,\, p_S(\vec{k}) = \braket{\Psi_{\vec{k}}|\mathcal{P}|\Psi_{\vec{k}}}, \\
    \ket{\Psi^{\bar{S}}_{\vec{k}}} &= \frac{1}{\sqrt{p_{\bar{S}}(\vec{k})}} (1-\mathcal{P}) \ket{\Psi_{\vec{k}}}, \,\, p_{\bar{S}}(\vec{k}) = 1-p_S(\vec{k}).
\end{split}\end{align}
Thinking of $\ket{\Psi^S_{\vec{k}}}$ and $\ket{\Psi^{\bar{S}}_{\vec{k}}}$ as Bloch bands, we can each associate with a separate quantum metric, just like in \equref{FullQuantumMetric}, which we will denote by $g^S$ and $g^{\bar{S}}$, respectively. Inserting the expansion (\ref{SplittinUpTheWavefunction}) into \equref{FullQuantumMetric}, we can relate the ``full'' quantum metric $g$ to the one of the projected subsystem and its complement as
\begin{equation}
    g_{ij} = p_S g^S_{ij} + p_{\bar{S}} g^{\bar{S}}_{ij} + \frac{\partial_i p_S \partial_j p_S}{4 p_S p_{\bar{S}}} + p_S p_{\bar{S}} \delta \mathcal{A}_i \delta \mathcal{A}_j. \label{SplittingUpTheeQuantumMetric}
\end{equation}
Here $\delta \mathcal{A}_i$ are the components of the differences of the Berry connections of the Bloch bands $\ket{\Psi^S_{\vec{k}}}$ and $\ket{\Psi^{\bar{S}}_{\vec{k}}}$, i.e., $\delta \mathcal{A}_i = \mathcal{A}^S_i - \mathcal{A}^{\bar{S}}_i$ where $\mathcal{A}^S_i = i \braket{\Psi^S_{\vec{k}}|\partial_{k_i} \Psi^S_{\vec{k}}}$ and $\mathcal{A}^{\bar{S}}_i = i \braket{\Psi^{\bar{S}}_{\vec{k}}|\partial_{k_i} \Psi^{\bar{S}}_{\vec{k}}}$. We note that while $\mathcal{A}^S_i$ and $\mathcal{A}^{\bar{S}}_i$ are individually not invariant under the gauge transformation (\ref{GaugeTransformationBlochStates}), their difference $\delta \mathcal{A}_i$ is, as required by consistency of the expression in \equref{SplittingUpTheeQuantumMetric}.

For a system like TBG with $C_{2z}\Theta$ symmetry, the last term in \equref{SplittingUpTheeQuantumMetric} is zero; it will more generally vanish (be negligible) if the Bloch band $\ket{\Psi^S_{\vec{k}}}$ associated with the top layer has (approximately) the same Berry connection as the one associated with the complement $\ket{\Psi^{\bar{S}}_{\vec{k}}}$. If the weight $p_S$ is approximately independent of momentum, as found for TBG, see \appref{ProjectionsInTBG}, we can further neglect the second term in \equref{SplittingUpTheeQuantumMetric}, leading to $g_{ij} \approx p^S g^S_{ij} + p^{\bar{S}} g^{\bar{S}}_{ij}$. 

Modulo the projection onto the $\vec{g}_0=0$ sector, we can reconstruct the full quantum metric in TBG. Since then $p_S \approx 1/2$, see \figref{fig:app_layer}, we have $g_{ij} \approx  (g^S_{ij} + g^{\bar{S}}_{ij})/2$. As long as no significant displacement field is applied, we can use the $C_{2x}$ symmetry of TBG to relate the two layers to one another, yielding $g^{\bar{S}}_{ij}=g^S_{C_{2x}i,C_{2x}j}(C_{2x}\vec{k})$. As such, we can obtain the full quantum metric from $g_{ij}(\vec{k}) \approx (g^S_{ij}(\vec{k}) + g^S_{C_{2x}i,C_{2x}j}(C_{2x}\vec{k}))/2$. Since the scanned paths are $C_{2x}$ symmetric, we can reconstruct the full quantum metric of TBG in the  $\vec{g}_0=0$ sector.

\subsection{Other degrees of freedom}\label{OtherDegreesOfFreedom}
Let us next discuss the impact of additional internal degrees of freedom beyond sublattice, such as spin. So while $\ket{\Psi^S_{\vec{k}}}$ might have more than the two components from the A and B sublattice, our procedure only provides access to
\begin{equation}
    \vec{n}(\vec{k}) := \braket{\Psi^S_{\vec{k}}|\vec{\sigma}\otimes \mathbbm{1}_{\bar{\sigma}}|\Psi^S_{\vec{k}}},
\end{equation}
where $\mathbbm{1}_{\bar{\sigma}}$ stands for the identity matrix in all remaining subspaces and, as in the main text, $\vec{\sigma}$ are Pauli matrices in sublattice space. Let us now define the reduced density matrix 
\begin{equation}
    P^S_r(\vec{k}) := \text{tr}_{\bar{\sigma}}[P^S(\vec{k})], \quad P^S(\vec{k}) = \ket{\Psi^S_{\vec{k}}}\bra{\Psi^S_{\vec{k}}},
\end{equation}
where $\text{tr}_{\bar{\sigma}}[\dots ]$ denotes the trace over all degrees of freedom other than sublattice (e.g., spin). Note that $P^S_r(\vec{k})$ is in general not a projector anymore, i.e., $[P^S_r(\vec{k})]^2 \neq P^S_r(\vec{k})$. Importantly, it now holds $\vec{n} =  \text{tr}[\vec{\sigma} P_r]$, and, thus,
\begin{equation}
    \frac{1}{4} (\partial_i \vec{n})\cdot (\partial_j \vec{n}) = \text{Re} \left[ \text{tr} \left[ P^S_r \partial_i P^S_r \partial_j P^S_r \right] \right] =:g^r_{ij}.
\end{equation}
This shows that our procedure of using \equref{gthetatheta} of the main text with the $\vec{n}$ obtained from the tunneling measurements in general actually extracts the ``reduced quantum metric'' $g^r$ in the sense that $P$ in \equref{FullQuantumMetric} is replaced by $P_r$. This also makes intuitive sense since the protocol is only sensitive to the momentum dependence of the Bloch state with respect to the sublattice degree of freedom and essentially ``blind'' to anything else. 

If our system of interest carries only spin on top of the sublattice degrees of freedom, it crucially depends on whether there is (non-collinear) spin-orbit coupling (like Rashba spin-orbit coupling) or not: if there is, the spin will wind in momentum space and, moreover, be entangled with the sublattice degree of freedom. The latter leads to $[P^S_r(\vec{k})]^2 \neq P^S_r(\vec{k})$ and $\vec{n}^2 < 1$. Furthermore, our current procedure, will only provide access to ``winding in sublattice space'', as formally captured by $g^r\neq g^S$. If spin-orbit coupling is negligible, as is the case in TBG, spin and sublattice are not entangled. Picking one spin flavor for concreteness [the presence of two is just encoded in the overall irrelevant prefactor $\mathcal{N}$, e.g., in \equref{eq:gamma_def}], the wave function will have the form $\ket{\Psi^S_{\vec{k}}} = \ket{\phi(\vec{k})}\otimes \ket{\uparrow}$. Then the trace over spin will lead to $P^S_r = \ket{\phi(\vec{k})} \bra{\phi(\vec{k})}$, which yields
\begin{equation}
    g^S = \text{Re} \left[ \text{tr} \left[ P^S \partial_i P^S \partial_j P^S \right] \right] = \text{Re} \left[ \text{tr} \left[ P_r^S \partial_i P_r^S \partial_j P_r^S \right] \right] = g^S_r.
\end{equation}

\section{Effect of projection in TBG}\label{ProjectionsInTBG}

The protocol that we developed in this work gives access to the sample wave-function projected in some degrees of freedom (e.g., layer) and traced out in others (e.g., spin). In this appendix, we discuss the effects of these projections for TBG flat bands. 

We can represent a given TBG state with a given spin and valley as
\begin{equation}
\ket{\psi_n(\vec{k})}
  = \sum_{\ell,\vec{g},\sigma} c^{(n)}_{\vec{g}\ell\sigma}(\vec{k})\,
    \ket{\vec{g},\ell,\sigma}_{\vec{k}},
  \qquad
  \ket{\vec{g},\ell,\sigma}_{\vec{k}}
  \equiv \bigl|\,\vec{k}-\kappa_\ell-\vec{g};\,\ell,\sigma\,\bigr\rangle .
  \label{eq:state}
\end{equation}
where $l=t,b$ indexes the top and bottom layers, such as $\kappa_b=\kappa$, $\kappa_t=\kappa'$ corresponds to the Dirac points used in the moiré reconstruction. Here $\vec{g}$ are the vectors of the moiré reciprocal lattice $\Lambda$ and $\sigma$ is the sublattice index. Furthermore, the states are normalized, so $\sum_{\ell,\vec{g}, \sigma} |c^{(n)}_{\vec{g}\ell\sigma}(\vec{k})|^2 =1 $.

\begin{figure}
    \centering
    \includegraphics[width=0.5\linewidth]{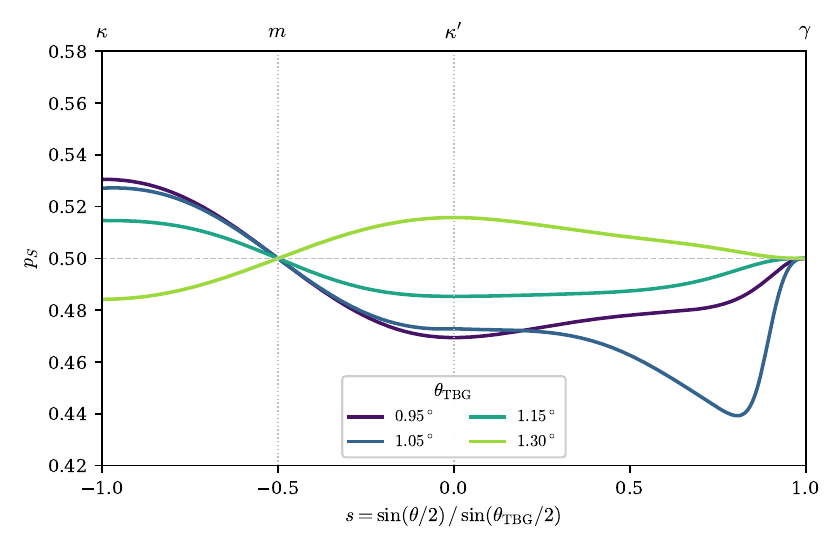}
    \caption{Layer weight for TBG lowest energy flat band $p_S(\vec{k)}$ along the scanned path for TBG flat bands }
    \label{fig:app_layer}
\end{figure}

Due to the fact that the tunneling between the tip and the sample happens mostly through the sample's top layer, our protocol accesses a layer-projected version of a given state. In \figref{fig:app_layer}, we can see that for a given TBG flat band $n$ (here the lowest energy one), the top layer projection $\mathcal{P}_t = \sum_{\vec{g}\in\Lambda}\sum_{\sigma=A,B} \bigl|\vec{g},t,\sigma\bigr\rangle\bigl\langle \vec{g},t,\sigma\bigr|$ gives approximately a constant weight $p^{(n)}_{S}(\vec{k}) = \bigl\langle\psi_n(\vec{k})\bigr|\mathcal{P}_t\bigl|\psi_n(\vec{k})\bigr\rangle$ along the scanned path. We also note that the $C_{2x}$ invariant point $m$ and $\gamma$ the top and bottom layer are equivalent and have exactly $p_S^{(n)} = \frac{1}{2}$.

\begin{figure}
    \centering
    \includegraphics[width=0.5\linewidth]{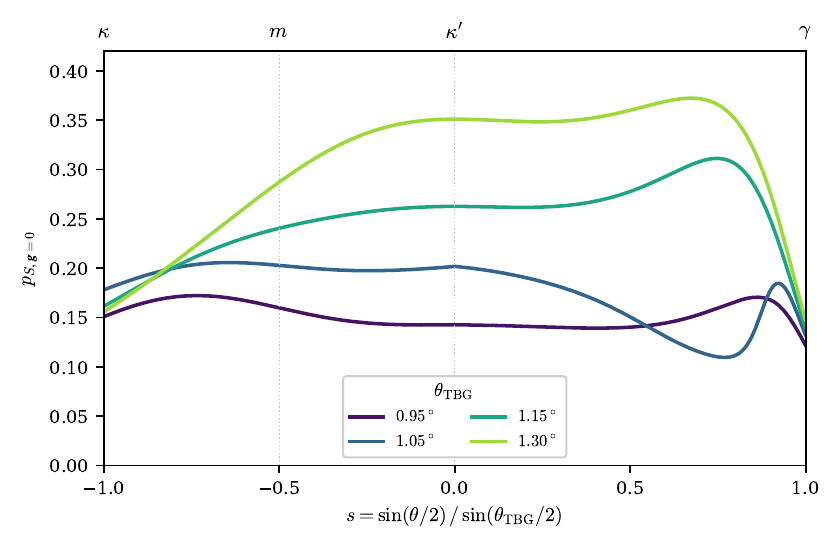}
    \caption{Weight of the top layer $\vec{g}=0$ component along the scanned path for TBG lowest energy flat band. Here we show a few angles near the ``magic angle''.}
    \label{fig:app_weight}
\end{figure}

Furthermore, if one scans a path that stays within one moiré Brillouin zone, the probed momentum corresponds to a displacement from the Dirac point $K'$ that is smaller than any $\vec{g}$; hence, the only $\vec{g}$ that can satisfy the tunneling condition is $\vec{g}=0$. Therefore, we only probe that component of a given TBG state, i.e., we must take the projection $\mathcal{P}_{t,\bm{0}} = \sum_{\sigma=A,B}  \bigl|t,\bm{0},\sigma\bigr\rangle\bigl\langle t,\bm{0},\sigma\bigr|$. In \figref{fig:app_weight}, we plot $p^{(n)}_{S,\vec{g}=0}(\vec{k}) = \bigl\langle\psi_n(\vec{k})\bigr|\mathcal{P}_{t,\bm{0}}\bigl|\psi_n(\vec{k})\bigr\rangle$. We can see that along this path, the weight is approximately constant, specially for TBG twist angles $\theta_{TBG}$ near the ``magic angle'' $1.05^\circ$.

\section{Consistency checks through Dirac cones and other points with known behavior}\label{ConsistencyCheckAppendix}

Let us consider two points that are equidistant from a Dirac cone and sufficiently close to it such that linearization of the effective Hamiltonian in momentum deviations from the cone is applicable. Let us denote the associated two angles by $\theta_{1,2}$. We then know that $n_\chi \left(\theta_1\right) = - n_\chi\left(\theta_2\right)$. We also have from \equref{nChiNz} that
\begin{equation}
    n_\chi = \frac{2\Gamma_1 - p_S\mathcal{N}(w_0^2+w_1^2)}{2 p_S\mathcal{N} w_0 w_1}.
\end{equation}
Hence, we can compare the two measurements to obtain
\begin{equation}
    \Lambda \equiv \Gamma_1(\theta_1)+\Gamma_1(\theta_2) = p_S\mathcal{N}(w_0^2+w_1^2),
\end{equation}
and therefore extract $p_S \mathcal{N} (w_0^2 + w_1^2)$ without further support from theoretical models.

Furthermore, $r \equiv \frac{w_0}{w_1}$ can be obtained through the traditional QTM \cite{wei2025dirac}. We are then able to write
\begin{equation}
\begin{split}
    p_S\mathcal{N}\,w_0w_1 &= \Lambda\,\frac{r}{1+r^2}, \\
    p_S\mathcal{N}\,(w_0^2-w_1^2) &= \Lambda\,\frac{r^2-1}{r^2+1}.
\end{split}
\end{equation}
Now, the Bloch components that are directly measured can be expressed through experimentally accessible constants and $\Gamma_{1,2}$
\begin{equation}
\begin{split}
    n_\chi &= \frac{1+r^2}{2r}\left(\frac{2\Gamma_1}{\Lambda}-1\right), \\
    n_z &= \frac{1+r^2}{r^2-1}\,\frac{2\Gamma_2}{\Lambda}.
\end{split}
\end{equation}

\end{appendix}

\end{document}